\documentclass[mathleft]{an}
\usepackage{graphicx}
\usepackage{times}
\newcommand{\procspie}{Proc.~SPIE}

\usepackage{natbib}
\usepackage{url}
\bibpunct{(}{)}{;}{a}{}{,}

\def\kms{$\mathrm{km\, s^{-1}}$}
\newcommand{\logg}{\ensuremath{\log g}}

\begin{document}

\Pagespan{1}{}
\Yearpublication{2012}%
\Yearsubmission{2012}%
\Month{01}%
\Volume{999}%
\Issue{999}%

\title{Velocity and abundance precisions for future high-resolution spectroscopic surveys:
a study for 4MOST}

\author{%
Elisabetta Caffau\inst{1,2}\fnmsep\thanks{Corresponding author: \email{ecaffau@lsw.uni-heidelberg.de}, Gliese Fellow\newline}, 
Andreas Koch\inst{1}, Luca Sbordone\inst{1,2}, Paola Sartoretti\inst{2}, 
Camilla J. Hansen\inst{1}, 
Fr\'ed\'eric Royer\inst{2}, Nicolas Leclerc\inst{2}, Piercarlo Bonifacio\inst{2},
Norbert Christlieb\inst{1}, Hans-G\"unter Ludwig\inst{1,2}, Eva K. Grebel\inst{3},
Roelof S. de Jong\inst{4}, Cristina Chiappini\inst{4}, Jakob Walcher\inst{4}, 
Shan Mignot\inst{2}, Sofia Feltzing\inst{5}, Mathieu Cohen\inst{2}, Ivan Minchev\inst{4}, 
Amina Helmi\inst{6}, Tilman Piffl\inst{4}, Eric Depagne\inst{4},
\and Olivier Schnurr\inst{4}
}
\titlerunning{Performance tests for 4\,m class telescopes' spectrographs}
\authorrunning{E. Caffau}
\institute{
Landessternwarte, Zentrum f\"ur Astronomie der Universit\"at Heidelberg, K\"onigstuhl 12, Heidelberg, Germany
\and 
GEPI, Observatoire de Paris, CNRS, Univ. Paris Diderot, 5 place Jules Janssen, 92195 Meudon Cedex, France
\and
Astronomisches Rechen-Institut, Zentrum f\"ur Astronomie der Universit\"at Heidelberg, M\"onchhofstr.\ 12-14, 69120 Heidelberg, Germany
\and 
Leibniz-Institut f\"ur Astrophysik Potsdam, An der Sternwarte 16, 
D-14482 Potsdam, Germany
\and
Lund Observatory, Department of astronomy and theoretical physics, Box 43, SE-221 00 Lund,
Sweden
\and
Kapteyn Astronomical Institute, University of Groningen, PO Box 800, 9700 AV Groningen, the Netherlands
}

\received{01 Jan 2012}
\accepted{01 Jan 2012}
\publonline{later}


\abstract{In preparation for future, large-scale, multi-object, 
high-resolution spectroscopic surveys of the Galaxy, we 
present a series of tests of the precision in radial velocity and chemical abundances
that any such project  can achieve at a 4\,m class telescope. 
We briefly discuss a number of science cases that aim at studying the chemo-dynamical history of the major Galactic components (bulge, thin and thick disks, and 
halo) -- either as a follow-up to the Gaia mission or on their own merits.
Based on a large grid of synthetic spectra that cover the full range in stellar parameters of typical survey targets,
we devise an optimal wavelength range and argue for a moderately high-resolution spectrograph. 
As a result, the kinematic precision is not limited by any of these factors, but will practically only suffer from systematic effects, 
easily reaching uncertainties $<$1\kms. 
Under realistic survey conditions 
(namely, considering stars brighter than $r=16$\,mag with reasonable exposure times) 
we prefer an  ideal resolving power of $R\sim$20000 on average, for an overall wavelength range (with a common two-arm spectrograph design) of 
[395;456.5] nm and [587;673] nm. 
We show for the first time on a general basis that it is possible to measure chemical abundance ratios to better than 0.1\,dex 
for many species 
(Fe, Mg, Si, Ca, Ti, Na, Al, V, Cr, Mn, Co, Ni, Y, Ba, Nd, Eu) 
and to an accuracy of about 0.2\,dex for other species such as Zr, La, and Sr.  
While our feasibility study was explicitly carried out for the 4MOST facility, the results can  be readily applied to and used for any other conceptual design study for 
high-resolution spectrographs. 
}
 
\maketitle
\sloppy
 

\section{Introduction}

The formation and evolution of a galaxy like the Milky Way (MW) still poses many puzzles to modern astrophysics.
The Gaia satellite, expected to be launched  in 2013, will make an important step 
towards addressing many of those questions pertaining 
to Galactic evolution (Munari \& Castelli 2000; Munari 2001; Perryman et al. 2001; 
Katz et al. 2004; Bailer-Jones 2002; Lindegren 2010). 
In the first place, Gaia will deliver accurate  stellar distances from parallaxes  and, coupled with proper motion measurements, 
transverse velocities  for more than 10$^9$ stars as faint as V$\sim$20\,mag. 
Radial velocities can also be obtained from Gaia's spectrograph, although this will be limited 
to approximately $1.5\times 10^8$ stars between 12 and 16\,mag. 
An important drawback is that the radial velocity spectrometer aboard Gaia only covers the 
\ion{Ca}{ii}  
triplet region, 847--874\,nm \citep{2004MNRAS.354.1223K},
so that only for a fraction of the brighter stars, limited information about the chemical 
composition will be obtained.  Moreover, it will not be possible to measure detailed 
element abundance ratios for a wide range of elements with Gaia.

While Gaia will provide a three-dimensional map of up to a billion stars in the MW 
(about 1\% of the total number of stars) and thus revolutionise the understanding of our Galaxy,  
it is not sufficient in itself to address a number of important questions in Galactic evolution.

Gaia will benefit hugely from additional,  ground-based spectroscopic information 
(detailed chemical abundances and precise line-of-sight velocities) for a broad range 
of stellar types, thus probing the various populations of the MW.  
Large survey missions that are currently designed as follow-ups to Gaia, where 
it lacks spectroscopic sensitivity, come with their own merits 
and allow us to gain ever deeper knowledge about the formation and chemo-dynamical evolution of the MW.

Present day multi-fibre facilities\footnote{For a comprehensive overview of currently commissioned 
and planned projects see \url{http://www.ing.iac.es/weave/moslinks.html}.} 
such as the European Southern Observatory/Very Large Telescope (ESO/VLT) multi-object facility 
FLAMES (140 fibres; Pasquini et al. 2002), Australian Astronomical Observatory's (AAO) 
AAOmega+2dF instrument ($\sim$400 fibres;  Sharp et al. 2006), 
Hectoechelle (300 fibres; Szentgyorgyi et al. 2011)
and Hectospec (300 fibres; Fabricant et al. 2005) on the MMT,
MIKE-fibers (128 ``red'' and 128 ``blue'' fibres; 
Bernstein et al. 2003) on the Magellan telescope, 
or the 6dF instrument (150 fibres, Watson et al. 1998) on the Schmidt telescope 
of the AAO, currently engaged in the RAdial Velocity Experiment (see Steinmetz et al. 2006), 
can provide large numbers of spectra and reach intermediate resolutions
($R\la $20\,000 for FLAMES and $R\la $ 8\,000 for AAOmega and RAVE), 
but, as of yet, there is no fully operational, dedicated, full-time survey of all Galactic 
components in high-resolution mode, in the optical. A first step in this direction
is the public Gaia-ESO Survey (Gilmore et al. 2012), which started on Dec. 31, 
2011 using FLAMES in high-resolution modes to assess the chemo-dynamics of some 10$^5$ Galactic field and cluster stars.
The Apache Point Observatory Galactic Evolution Experiment, APOGEE at 2.5\,m APO, Eisenstein et al. (2011),
observes in the near infra-red, in the H band at a resolving power of about 30\,000.

In the era of Gaia, a considerable fraction of the astronomical community is giving thought on how to best take 
advantage of the information flow that  Gaia will bring along. 
Much effort is thus devoted to conceive large surveys to complement 
and complete the radial velocity and chemical pattern information in the MW.
For details on ongoing surveys and instruments the interested reader is referred to 
RAVE, Steinmetz et al. (2006);
Sloan Extension for Galactic Understanding and Exploration, SEGUE, Yanny et al. (2009); and APOGEE.
Details on some studies on instrumentation are documented for  
4MOST (at the 4.1\,m Visible and Infrared Survey Telescope for Astronomy, VISTA) in de Jong (2011), de Jong et al. (2012a,b); 
GYES at the  3.6\,m CFHT, Bonifacio et al. (2011);
High Efficiency and Resolution Multi-Element Spectrograph, HERMES, at the 4\,m Anglo-Australian Telescope, AAT, Barden et al. (2010); 
MOONS, Multi-Object Optical and Near-infrared Spectrograph, at the 8.2\,m VLT, Cirasuolo et al. (2011); and 
WEAVE at the 4.2\,m William Herschel Telescope, WHT, Balcells et al. (2010). 
In this context, a detailed study of the required performance and characteristics of such spectrographs is of prime importance. 

In general, considerations for these fibre-fed spectroscopic survey facilities are 
driven by  a large field of view to cover large fractions of the sky, by  
a generous multiplexing capability and high enough spectral resolving power  
to detect kinematic and chemical substructures throughout all major components of the MW 
(thin and thick disks, bulge, and halo), and by a broad wavelength coverage to ensure a high measurement precision and the detectability of 
as large a number of elements as possible and to be able to derive the stellar parameters and the chemical
composition of the stars at best.

What follows refers especially
to our effort in understanding and defining the characteristics that a typical survey-spectrograph 
should have in order to achieve the precision necessary to fulfil typical 
science cases (see Sect.~2) that ground-based Gaia follow-up surveys endeavour to pursue. 
While the present tests have been explicitly run as a conceptual design study for the 4MOST\footnote{
4MOST, the {\em 4\,metre Multi-Object Spectroscopic Telescope},  is a high-multiplex, wide-field  (3\,degree$^2$
field-of-view) fibre-fed spectrograph
system in its phase A study. It is intended for the ESO/VISTA 4.1-m telescope
to work on a dedicated 5-years survey base, starting in 2019. 
The spectrograph is designed to provide simultaneously at least 1500 (with a goal of 3\,000) low- 
(R$>$5\,000) and high-resolution (R$\sim$20\,000) spectra.   
We note that the definition of ``high'' and ``low'' resolution is often 
somewhat arbitrary in the literature. To distinguish the two operation modes of 4MOST we simply call them
high- and low-resolution in the following.
In the end, it is envisioned to  obtain tens of millions of spectra at low resolution 
and more than 1 million spectra at high resolution. 
The bulk of the high resolution observations of
4MOST will extend down to an $r$-band magnitude of 16\,mag at a signal-to-noise ratio 
of $>$ 150 \AA$^{-1}$
(which is translated in a S/N ratio of $>$ 44 pixel$^{-1}$ at the centre of the blue range
with a resolving power of 20\,000 and a sampling of 2.5)
in less than 2-hour exposures. 
This will ultimately allow one to achieve the science goals outlined in Section~2.
For details on 4MOST see de Jong (2011); de Jong et al. (2012a, 2012b).
} 
facility, our methods and results are generalised and can be readily applied  for phase A studies of
any other, potential  spectrograph projects in the future.
This paper is dedicated to the {\em high}-resolution mode of the 4MOST  survey (and others). 
An analogous study of the low-resolution capabilities of such instruments will be presented in a future paper. 
In particular, here we elaborate on our efforts to determine the optimal wavelength range and the 
resolution of any such spectrograph, given the signal-to-noise ($S/N$) ratios necessary to achieve the
accuracy requested for typical Galactic science cases.

This paper is organised as follows: 
in \textsection2 we give an overview of typical science cases that drive the design of ground-based survey strategies and the requirements these pose to a survey instrument. 
In \textsection3 we discuss the target range and spectral types that need to be considered in the design studies, while \textsection4 describes the grid of synthetic  spectra 
we employ for our study. \textsection5 deals with our tests on the technical requirements to the spectrograph based on the science questions. We close with the discussion of our 
design study in \textsection6.  

%
%
%
\section{General science questions}
The mechanisms of the formation and evolution of the major MW components are encoded 
in the location, kinematics, and chemistry of their stars: physical conditions of star 
formation suggest that most of the stellar material is produced in a finite number of 
large clumps,  characterised by some amount of chemical homogeneity. Various dynamical 
processes that act upon these stellar aggregates (spiral arms, bar, interactions or 
mergers with satellites) can erase the kinematic imprint of objects born together. Also,  
lighter clusters will simply disperse due to the original internal velocity dispersion 
and general interaction with the gravitational field of the MW.  

Detailed stellar chemical compositions can then be used as genetic markers to unravel the lineage of common field stars. 
In this context, prime science drivers are inevitably the 
chemical tagging of any chemo-dynamical substructures (Freeman \& Bland-Hawthorn 2002; Bland-Hawthorn, Krumholz \&
Freeman 2010), 
including abundance gradients and their temporal variations (e.g., Chiappini et al. 2001), 
 and searches for (extremely) metal-poor stars in the halo and bulge such as to characterise 
the first stellar generations in the Universe, thereby constraining galaxy formation at the earliest times (e.g., Beers \& Christlieb 2005, Karlsson et al. 2011).

For turn-off stars and sub-giants it will be possible to derive their ages directly based
on the parallaxes measured by Gaia, however, for stars on the main-sequence and
giant stars no age information is available. 
The age-metallicity and age-velocity relations for a stellar population represent strong constraints on any chemo-dynamical model. 
As ages are only available for turn-off stars this 
information will be limited to the more nearby stars (about 1\,kpc). 
In practice it will 
still be feasible to determine rough ages using ``chemical clocks'' (such as $\alpha$ 
and neutron capture elements; e.g., Tinsley 1976, Matteucci 2003). This can be done 
with ease also at such large distances as the outer disk and the Galactic bulge. 

\subsection{Thin and thick disks}
Despite recent, major progress in understanding disk formation in cosmological 
simulations, many discrepancies remain between the predictions of simulations and the 
observed properties of disk galaxies (e.g., Scannapieco et al. 2009, Piontek \& Steinmetz 2011). 
In fact, mergers may not necessarily be the dominant process in the 
formation of disks, but internal evolution processes such as radial migration may also play an important 
role in shaping disk galaxies (e.g., Sch\"onrich \& Binney 2009,  Minchev et al. 2012).

Recent advances in high resolution spectroscopy have shown that a kinematic separation
between a thick 
and a thin disk is, in principle, statistically, feasible. The two populations show distinct trends in chemical
abundances, but when the statistical decomposition is used there is naturally an overlap
(see, e.g., Fuhrmann 2008; Bensby et al. 2005; Reddy et al. 2006; Ruchti et al 2011). 
It is potentially possible to identify the different abundance patterns
as the result of different star formation histories (Chiappini 2011) but they can also be 
explained as the result of internal redistribution of the stars in the stellar disk (e.g.,
Sch\"onrich and Binney 2009, cf. Minchev et al. 2012).

However, an important criticism of current observational chemo-dynamical 
studies of the stellar disk is that the samples are confined to the 
immediate solar neighbourhood (e.g., Steinmetz et al. 2006), 
and that the separation into thin and thick disk stars is solely based on the stellar kinematics, while radial 
migration of stars (Minchev \& Famaey 2010) or accretion (e.g., Williams et al. 2011 and references therein) 
could bring to the solar neighbourhood stars that were born elsewhere contaminating the ``local'' samples 
(e.g., Grenon 1999; Haywood 2008). 
Stars of a given age would then show a spread in metallicity equal to 
the radial spread in the metallicities of the interstellar medium at their time of birth. 
Radial migration effects are thus manifested in the magnitude and evolution of the radial abundance gradients in the disk.

Chemo-dynamical models provide very different predictions for several key observables (e.g., age-metallicity and age-velocity relations, 
abundance patterns such as [X/Fe] vs. [Fe/H] and the scatter in these relations) at different galactocentric distances and heights from the plane, 
depending on the details of radial migration and chemical evolution model. Moreover, stars from accreted satellites could dominate the outer disk, 
while any population formed in-situ will be more important closer to the Galactic plane (Villalobos \& Helmi 2009). In particular, farther out in the disks the predictions 
are very model-dependent as in those regions additional mechanisms are at play, such as star formation thresholds, accretion, heating and stripping by dark matter halos. 
Therefore, full constraints on the formation of this component can only be obtained through joint kinematic and abundance surveys covering the entire Galactic disk to a high accuracy.
In this context, it has been recently demonstrated that the eccentricity distribution of orbits 
in the thick disk can be linked to the 
accretion vs. heating vs. migration scenarios (Sales et al. 2009; Wilson et al. 2010; Dierickx et al. 2010). 
To make significant progress in the entire field, it is necessary to map the density of stars 
in a multi-dimensional space, namely comprising age, metallicity, position and the 3-dimensional velocities.

\subsection{The bulge}
Two main formation scenarios have been proposed to explain bulges of spiral galaxies (Kormendy \& Kennicutt 2004). In classical bulges, most stars are formed during an early phase of intensive 
star formation following either the collapse of a proto-galactic cloud or a phase of intense merger/accretion events as predicted by cold dark matter simulations. On the other hand, boxy or peanut-shaped, 
pseudo-bulges are a consequence of secular evolution driven by dynamical instabilities in the inner disk, such as the buckling of a bar (Combes et al. 1990). The different scenarios predict
 different timescales of formation, and hence different chemo-dynamical histories for bulges.
 
The formation of the Galactic bulge appears to have happened through a
combination of processes:
high-resolution spectroscopic studies of giants along low 
extinction windows (Zoccali et al. 2006; 2008; \citealt{hill11,gonzalez}) 
have revealed mostly old $\alpha$-enhanced stars, compatible with a classical bulge (cf. Bensby et al. 2011). 
On the other hand, its boxy shape, revealed in the near-infrared   
\citep{dwek,binney},
and its kinematics (e.g., Howard et al. 2008, 2009; Shen et al. 2010) suggest that our Galaxy harbours a pseudo-bulge. 
Likewise, the co-existence of  two red clump populations 
in particular fields toward the Galactic bulge (McWilliam \& Zoccali 2010) 
argue for an X-shaped, double-funnel shape of the Galactic bulge/bar. 

The power of combining chemistry with kinematic studies in the bulge is illustrated by the recent work of Babusiaux et al. (2010), who found  two distinct populations in the bulge, 
namely a metal-rich, $\alpha$-poor population with bar-like kinematics, and a metal-poor, $\alpha$-enhanced population that shows rather spheroidal or thick-disk like properties. 
The only way to clarify the critical questions on the current structure of the Galactic bulge, its 
formation and the transition to other Galactic components
is to map the distribution of elemental abundances to high precision and to couple these to accurate kinematics
over a wide range of bulge locations, thereby unveiling the mix of the different populations 
(bar, spheroidals) at different distances from the plane (see also Johnson et al. 2011). 
This will also enable studies of the earliest phases of the Galactic chemical 
enrichment, properly sampling the metal-poor tail of the bulge metallicity 
distribution function.  It is conjectured that 
the first stars formed are either found in the bulge (Tumlinson 2010) or the halo (Sect.~2.3). 
Since metal-poor stars are very rare in the bulge this dissection of the growth 
and evolution of the bulge requires surveys of large numbers of stars (Kunder et al. 2012). 

\subsection{Galactic halo}
The Galactic halo contains the oldest and most metal-poor stars currently known in our Galaxy. 
Their metallicity distribution and the abundances of the chemical elements (chemical abundances)
accessible to high-resolution surveys contain unique imprints of the physical processes 
that took place from the formation of the first stars to the build-up of our Galaxy
(Karlsson et al. 2011).
In particular, 
stars below [Fe/H] $\sim -2.5$\,dex 
can provide indirect information on the  star formation processes in metal-poor environments, 
the properties of the first generation of stars (e.g., their initial mass function or rotation characteristics), 
and probe the presence of a critical metallicity for star formation 
(Caffau et al. 2011a; Klessen et al. 2012), as well as any dependence on environment 
(as may be the case for different halo building blocks).

However, only three stars at [Fe/H] 
$\la-$5.0\,dex are known to date. 
The samples available today are also mostly dominated by stars associated with the {\em inner} halo, 
since only a small fraction of the samples reach Galactocentric distances larger than 15 kpc, 
where the outer halo population dominates (Carollo et al. 2007).	
Extrapolating the halo metallicity distribution of Sch\"orck et al. (2009), one can expect to unearth 
at least 10\,000 objects below [Fe/H]=$-2$\,dex under a typical survey strategy (at some 10$^5$ targets), 
still yielding several hundreds objects below $-$4\,dex and a few tens below $-$5\,dex. 

We also note  that Nissen \& Schuster (2010)
claimed the distinction of two groups of stars   in the [$\alpha$/Fe] versus [Fe/H] 
abundance ratio space. These groups can be
kinematically associated with the inner and outer halo, respectively. 
Future  high-resolution spectroscopy halo surveys will extend such studies to farther distances, 
using distances and proper motions obtained by Gaia, and applying it to a higher 
dimension abundance ratio space -- provided a high level of precision in the abundance measurements
can be achieved.
Likewise, 
chemical tagging will aid to identify substructures 
(such as hundreds of streams of disrupted satellites) associated with past 
merger and accretion events, as well as characterise the properties of the progenitor 
``building blocks'', such as their chemical enrichment history, and the main nucleosynthetic channels that led to their present abundance patterns (Tolstoy et al. 2009; Koch 2009). 

\subsection{Nucleosynthesis}
Detailed chemical abundance distributions will not only serve to disentangle the formation of the Galactic components, but can also provide invaluable insight into the 
origin of the elements themselves. Here we briefly touch upon two examples that reflect the relevance of high-resolution measurements of light {\em and} heavy elements
for the study of nucleosynthetic processes; their actual measurability will be discussed in Sect.~5.2.2. 

According to the standard big bang nucleosynthesis model, lithium is produced in primordial
nucleosynthesis. Metal-poor ([M/H]$<-1.4$) dwarfs, formed from primordial material enriched by only a few supernovae,   
have mainly a constant Li abundance, the so called
Spite plateau \citep{spite82}.
Recent studies have shown a meltdown in this plateau (Sbordone et al. 2010)  
that is not fully  understood.
As stars evolve to the giant branch, their convective zone reaches
higher layers in the stellar atmosphere, and the fragile Li is destroyed by the exposure to higher temperatures.
Among the targets we plan to observe there will be such metal-poor stars,
that will allow us to enlarge the statistics on the stars on the Spite plateau and beyond.
It has been recently shown \citep{Mucciarelli1,Mucciarelli2}
that old, red giants below the RGB bump display a constant
lithium abundance that mirrors the Spite plateau at 
a lower level. This observation supports theoretical
computations that predict that those stars
have diluted their lithium content, but 
preserve memory of what their original lithium
content was. 
If confirmed by further, observations this would open up
the possibility of studying the original lithium
content in external galaxies in the Local Group, thanks
to the higher luminosity of giants with respect to dwarfs.

In spite of constituting only  small fraction of the  baryonic mass of the MW, $n$-capture elements
leave an important mark on stellar spectra over a large range in metallicity.
The actual formation processes of these heavy elements (e.g. the main $r$-process) remain to date one of 
the open questions in astrophysics. Detailed abundance information can allow us to understand 
the process and the conditions under which it works, the sites of the production of these
elements, {\em and} their production during  different epochs of the Galaxy.
For a complete review on this subject see Sneden et al. (2008).

\subsection{Implications for velocity and abundance requirements}

The ambitious goal of the 4MOST survey is to sample all major
nucleosynthetic channels namely, light elements, odd-Z elements (Na, Al),
$\alpha$ elements, Fe-peak elements, and heavy and light neutron-capture elements such as slow neutron-capture 
($s$-) process elements, and  rapid neutron-capture ($r$-) process elements.
This leads to the requirement that we
sample at least 15 different elements, as specified in Table\,\ref{range}.
This chemical study of a large sample of Galactic stars permits one to better understand our Galaxy.
A precise knowledge on Fe and $\alpha$-elements with kinematics would allow one to find and characterise different stellar populations in the MW.
A chemical map on heavy elements can help us understand their actual production processes and define the sites of their formation.
The precision in chemical abundances is driven by the need to distinguish different stellar populations; 
for instance, the two halo populations advocated by
\citet{NissenSchuster} are separated by 0.1--0.2\,dex
in [$\alpha$/Fe]. 
 
\begin{table*}[htb]
\begin{center}
\caption{Number of available lines in the proposed spectral wavelength ranges, separately for dwarfs and giant stars.
}
\begin{tabular}{|c | c c|| c c|}
\multicolumn{1}{c}{}  & \multicolumn{2}{c}{$[380;450]$\,nm domain} & \multicolumn{2}{c}{$[580;680]$\,nm domain} \\ \hline
Parameter & Element     & \# lines & Element     & \# lines \\
          &             &   (dwarf/giant)    &            (dwarf/giant) &  \\
\hline
 
$T_{\rm eff}$ & \ion{Fe}{i}         & 73/70              & Fe{\sc i}        & 96/107          \\
              &                   &                    & H$\alpha$   & 1            \\ \hline
$\log g$     & \ion{Fe}{ii}        &  12/7                     & Fe{\sc ii}        & 8/2      \\
\hline
[$\alpha$/Fe]& \ion{Ca}{i}         & 8/8               & Ca{\sc i}         & 14/14        \\ 
             & \ion{Mg}{i}         & 2/1                & Mg{\sc i}         & 1/2         \\
             & \ion{Si}{i}         & 3/2                & Si{\sc i}         & 13/12       \\
             & \ion{Ti}{i}         & 13/19              & Ti{\sc i}         & 8/27        \\
             & \ion{Ti}{ii}        & 10/26              & Ti{\sc ii}        & 0/1         \\ \hline
[X/Fe]    &                   &                    &  Li{\sc i}        & 1               \\
             & \ion{Na}{i}         & 1/0                & Na{\sc i}         & 4/3         \\
             & \ion{Al}{i}         & 2/1                & Al{\sc i}         & 1/2         \\
             & \ion{Sc}{i}         & 1/0                & Sc{\sc i}         & 1/1        \\
             & \ion{Sc}{ii}        & 3/4                & Sc{\sc ii}        & 2/6        \\
             & \ion{V}{i}          & 7/7                & V{\sc i}          & 9/27        \\
             & \ion{V}{ii}         & 1/1                & V{\sc ii}         & 0/0         \\
             & \ion{Cr}{i}         & 6/4                & Cr{\sc i}         & 1/2       \\
             & \ion{Cr}{ii}        & 1/1                & Cr{\sc ii}        & 0/0       \\
             & \ion{Mn}{i}         & 19/13              & Mn{\sc i}         & 2/3         \\
             & \ion{Mn}{ii}        & 0/1                & Mn{\sc i}         & 0/3         \\
             & \ion{Co}{i}         &  9/11              & Co{\sc i}         & 0/11        \\
             & \ion{Ni}{i}         & 3/3                & Ni{\sc i}         & 18/19       \\
             & \ion{Sr}{ii}        & 1/1                & Sr{\sc ii}        & 0/0        \\
             & \ion{Y}{ii}         & 3/1                & Y{\sc ii}         & 1/1         \\
             & \ion{Zr}{ii}        & 2/2                & Zr{\sc ii}        & 0/0        \\
             & \ion{Ba}{ii}        & 1/1                & Ba{\sc ii}        & 2/2        \\
             & \ion{La}{ii}        & 1/6                & La{\sc ii}        & 0/4        \\
             & \ion{Nd}{ii}        & 2/5                & Nd{\sc ii}        & 1/0        \\
             & \ion{Eu}{ii}        & 1/1                & Eu{\sc ii}        & 1/1        \\
             & CH          & G-band                  & &             \\
             & CN          & X-A band               & &   \\
\hline\end{tabular}
\label{range}
\end{center}
\end{table*}

According to the end-of-mission
performances assessed at the time of
the  Gaia Mission Critical Design Review (April 2011
\footnote{\url{http://www.rssd.esa.int/index.php?project=GAIA&page=Science_Performance}}),
a V=15\,mag G2V star will have an error on the parallax
of 24\,$\mu$as. Neglecting interstellar reddening and assuming
an absolute magnitude of $M_V$=4.4\,mag for such a star, the distance is
of the order of 1.3\,kpc. This means that for a proper motion of 1 mas yr$^{-1}$ the
error on the transverse 
velocity\footnote{Transverse velocity (VT), proper motion ($\mu$), 
and parallax ($\pi$) are linked through VT$= A_V {\mu\over\pi}$
\citep[][vol. 1 eqn. 1.2.20]{hipparcos}, $A_V = 4.74047$ [km yr s$^{-1}$],  
from which we
derive the error expression 
$\sigma_{\rm VT} = A_V{\sigma_\pi\over\pi}\left( 0.526+{\mu\over\pi}\right)$, 
since $\sigma_\pi$ and
$\sigma_\mu$ are correlated through $\sigma_\mu = 0.526\sigma_\pi$.
}
is $\sim$ 0.3\kms,
for 5 mas yr$^{-1}$ it is of $\sim $ 1.1\kms.  
The radial velocities derived from a survey built as
follow-up of Gaia on the ground should have a comparable precision.

%
%
%
%
\section{Targets}
Realistic estimates of spectrograph efficiency,
coupled with a 4\,m class telescope lead to a typical 
limiting magnitude  of $r<$16\,mag.
The parallaxes from Gaia in this regime will permit one to derive distances to within 10\% accuracy out to $\sim$10 kpc for red giants. 
The distance scale is, naturally, limited by the spectral type and, at V=16\,mag, F--G dwarf distances will be determined at the 10\% level out to  
maximally a few kpc (see also Fig.1 in de Jong 2011), which bears relevance to contriving the science drivers of follow-up surveys (Sect.~2). 
We concentrate on the case of dwarf stars because for them age measurements from isochrones will be possible.

Our science questions above imply that  more than 50\% of the targets observed with
the 4MOST high resolution mode will be
dwarfs with a metallicity in the range $-2.0<\left[{\rm Fe/H}\right]<+0.2$.
Moreover, since the Galactic bulge is of crucial interest to the spectroscopic surveys,  
a fraction of the targets will be metal-rich bulge giants. Yet 
another contribution to the samples will come from metal-poor giants as found in the Galactic halo.
Typical selection criteria for halo science cases restrict their targets  to stars with [Fe/H]$<-1$\,dex. 

A general concern  
is that studies of metal-poor giants require a predominantly blue  wavelength range, while 
spectroscopy of metal-rich giants benefits from a redder spectral coverage.
This is due to the fact that for wavelengths below about 450\,nm,
the spectrum of a metal-rich giant is an inextricable series of blends,
while for metal-poor giants only very few lines are strong enough to be detected for
wavelengths red-wards of about 500\,nm.

Chemical abundances for a large number
of elements for dwarf stars and metal-poor giants can generally be derived with high accuracy 
(see, e.g., Cayrel et al. 2004; Bonifacio et al. 2009), while this 
is far more challenging 
in the metal-rich giants of the bulge (Fulbright et al. 2007, Hill et al. 2011). 
Due to the high extinction, typically observable bulge stars 
are cool (${\rm T}_{\rm eff}\le 4200$\,K). For instance, 
\citet{hill11} observed stars in  Baade's window  with the  FLAMES/Giraffe instrument at the VLT; 
 their targets with an $r$-magnitude $< 16$\,mag have  effective temperatures 
in the range 4300--5300\,K (see also Johnson et al. 2011).

Surveys with 4\,m class telescopes 
are restricted to brighter objects, 
but one will also be able to cover regions with higher 
extinction than the low-extinction windows in the Bulge, such as Baade's and Plaut's fields so that
typical surveys will contain stars with ${\rm T}_{\rm eff}\le 4200$\,K. 
For these low-temperature stars, the TiO bands in the spectra become non-negligible,
and spectral synthesis taking into account the molecular bands
is mandatory for a detailed chemical analysis.
\citet{wylie09} list a number of iron lines that are free from the TiO bands and some of them fall in the spectral range 
foreseen for 4MOST (see also Fulbright et al. 2006).  
Overall, it is unlikely that the same number of chemical elements 
in stars of the bulge (e.g. Bensby et al. 2010; Ryde et al. 2010) 
can be measured, compared to, say, subsamples of the Galactic halo 
(e.g., Fulbright et al. 2007; Roederer \& Lawler 2012). 
 
%
%
%
\section{Simulated spectra}
We selected stellar parameters that represent typical
Galactic stars and cover the expected parameter space of common survey targets.
For dwarfs we selected two values of effective temperature
(6500\,K and 5500\,K) and two values of gravity (log\,$g$=3.0 and 4.0 in cgs units).
We computed synthetic spectra for
five metallicities ([M/H] = +0.5, 0.0, $-$1.0, $-$2.0, $-$3.0).
For giants we computed one single temperature (4500\,K), two 
values of gravity (2.0 and 1.0), and the same five values of metallicity.
The grid contains 30 models, 20 for dwarfs and 10 for the giants.

The model grid 
was computed by using 
ATLAS9 (Kurucz 2005) running under Linux  (Sbordone  2005).
For the opacity distribution functions we employed the Solar (ODF) and $\alpha$-enhanced 
(AODF) versions from \citet{castelli03}.
All metal-poor models ([M/H]=$-$1, $-$2, $-$3) 
have been computed with an enhancement in the $\alpha$-elements of 0.4\,dex, as is 
indeed seen in metal poor globular cluster and halo field stars (e.g., Barbuy 1983; Pritzl et al. 2005). 
The micro-turbulence of the models has been fixed at 1.0\kms, and the mixing-length 
parameter has been adopted as $\alpha$=1.25.

We computed the grid of synthetic spectra with the SYNTHE code 
(Kurucz 2005) running under Linux (Sbordone 2005),
in the range 370--700\,nm. 
The micro-turbulence in the synthesis was set to 1.5\kms.
No extra rotational velocity with respect to the V\,sin$i$=0.0\kms\
was included for all spectral tests. 
Zero-rotation is a reasonable approximation for a G-type dwarf, while 
problems may occur for early F-type stars, which can show
substantial rotational velocities (see Gray 2005).
In practice, we will derive V\,sin$i$ of the stars, either from
analysis of the width of the cross-correlation peak (Melo et al. 2001), or through
Fourier analysis of individual line profiles or of the cross-correlation function (Diaz et al. 2011). 
This parameter will then be used throughout the chemical analysis.
We foresee a standard line profile fitting for the abundance analysis.  
Thus, any extra-broadening of the lines due to rotation and ensuing progressive blending of features will not
strongly affect the precision in abundances due to our careful treatment of line blends during the fitting process. 
Only for the cases of high rotational velocities in excess of 
V\,sin$i \ga 20$ \kms,
the analysis will become difficult, but we estimate that the fraction of stars showing such properties will
be negligible in our target samples. 

Likewise, our analysis ignores the effects of micro- and macro-turbulence
on the uncertainty on the derived abundances.
These parameters are not physical quantities, but fudge-factors
introduced in the analysis based on one-dimensional, time-independent, 
stellar model atmospheres. 
The use of hydrodynamical model atmospheres removes the need for these
parameters (Steffen et al. 2009).
By the time of the data analysis of the 4MOST data, it can be expected that
hydrodynamical models will have reached a sufficient maturity to be routinely used. 
Since  micro- and macro-turbulence  depend on the known atmospheric parameters of the stars, 
an adequate calibration can then be reliably devised (see also Sect.~5.3).

The original  grid has a wavelength sampling
corresponding to a resolving power of 600,000.

These basic, synthetic spectra were then manipulated to provide the input for our tests 
of observations of resolution R$\sim$15000--25000 as follows.
In the first place we degraded the spectra to a resolving power of 250,000 -- 
about ten times  the value envisioned for a high-resolution spectrograph.  
Secondly, we normalised the spectral  flux to the corresponding $r$-band magnitude as 
$$F\, = \, 3631\times 10^{-23}\times 10^{-0.4 \, r}\times 
\left(\frac{hc\lambda}{6166^2\,f_{0}}\right)$$ in [erg\,s$^{-1}$\,cm$^{-1}$\,\AA$^{-1}$], 
where $f_{0}$ is the monochromatic flux at a reference wavelength of 6166\,\AA,
for two reference magnitudes of $r=14$ and 16\,mag. 
Finally, the spectra were 
resampled to an equispaced wavelength space with a step size in $\lambda$ of 0.5\,pm.  
These spectra have been used  to 
determine the spectrograph's design and spectral requirements to reach the science goals outlined above.

\begin{figure}[htb]
   \includegraphics[width=8cm,clip=true]{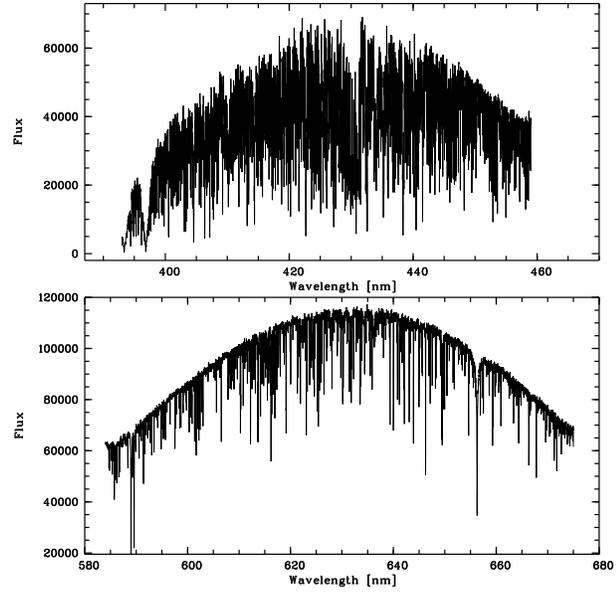} 
 \caption{Simulated spectra of the turn-off star 5500/4.0/0.0 of r=14 mag, observed in New Moon for 6000\,s.
Top panel shows the blue and lower panel the red arm.
The S/N ratio per pixel at the centre of the blue and red arms is 75 and 121, respectively.
}
    \label{spectra}
\end{figure}

The aforementioned synthetic spectra we processed through a throughput simulator 
that produces simulated observed spectra,
taking into account 
the characteristics of the targets 
(i.e., these very synthetic spectra), 
a transmission model of the Earth's atmosphere,  
the seeing,  the sky background, and 
a simple model of the instrument (see Sartoretti et al. 2012 for details). 
An example of a simulated, noise-added spectrum for  the case of a G-turn-off star
of solar metallicity is shown in Fig.\ref{spectra}.

The basic characteristics of the spectra computed with the simulator that we employ for the following studies 
are reported in Table\,\ref{design}.
 
\begin{table}[htb]     
\begin{center}       
\caption{Spectral characteristics of the simulated spectra.}   
\begin{tabular}{l | l | l }                                
\hline
 & Blue arm & Red arm\\
\hline                                                
\hline                                                
$\lambda _{\rm min}$  & 395\,nm       & 587\,nm \\
${\rm R}_{\rm min}$   & 18150         & 18635 \\
Sampling$ _{\rm min}$ & 2\,pixel/FWHM & 2\,pixel/FWHM \\
\hline
$\lambda _{\rm centre}$  & 425.8\,nm         & 630\,nm \\
${\rm R}_{\rm centre}$   & 19563             & 20\,000 \\
Sampling$ _{\rm centre}$ & 2.51\,pixel/FWHM & 2.5\,pixel/FWHM \\
\hline
$\lambda _{\rm max}$  & 456.5\,nm        & 673\,nm \\
${\rm R}_{\rm max}$   & 20976            & 21365 \\
Sampling$ _{\rm max}$ & 4.29\,pixel/FWHM & 4.22\,pixel/FWHM \\
\hline                                                
\end{tabular} 
\label{design}
\end{center}
\end{table} 

%
%
%
%
\section{High Resolution mode requirements}
\subsection{Wavelength range}

For our science case we need an as wide as possible range in wavelength.
We assume a spectrograph design with two arms.
If a mosaic of two detectors should be used to cover the needed area of the focal plane in each arm, 
the position of the inter-detector gap should also be carefully considered.

To fix the two wavelength ranges we tried to optimise the chemical output,
yielding abundances for the largest number of elements,
but giving more weight to the most interesting ones for our science goals.
The envisaged wavelength ranges are:
\begin{itemize}
\item
for the blue arm [392;456.5]\,nm or [395;459]\,nm, with the provision that
the CH G-band at [429;432]\,nm falls completely on one detector;
the two possibilities in range correspond to the \ion{Ca}{ii}-K
falling inside or outside the range, respectively;
\item
for the red arm [587;673]\,nm, with, if two detectors are necessary,
a gap around 635\,nm. 
\end{itemize}
The above ranges have been selected in order to optimise
the  stellar parameter determination  and to derive  
a maximally comprehensive set of chemical abundance patterns across the expected target range. 

An important question concerns the \ion{Ca}{ii}-K line (393.3663\,nm).
\ion{Ca}{ii}-K is a very strong line and, as such,  still detectable in the most metal-poor
objects. For instance,  in a star with $[{\rm M/H}]\approx -5$, this line still has an equivalent width 
(EW) of about 28\,pm, making it too strong to be a reliable indicator for the Ca abundance. 
At a spectral resolving power of $R=$20\,000,
this line is thus too strong to be useful for the large majority of the stars observed. 
The expected yield of ultra metal poor stars below [M/H]$< -5$ is very low 
(e.g., Christlieb et al. 2004; Frebel et al. 2005; Beers \& Christlieb 2005; 
Sch\"orck et al. 2009; Li et al. 2010; Caffau et al. 2011a) so that we chose 
to exclude the K-line from the covered wavelength range of the 
high-resolution mode. 
\ion{Ca}{ii}-K is a more important feature in  low-resolution studies 
(Beers et al. 1985; Beers \& Christlieb 2005), where lines that  strong   
 are the only ones that can still be detected at very low metallicities.

Following common practice, stellar parameters determinations generally rely on a combination 
of the indicators listed below (e.g., Edvardsson et al. 1993;  
Koch \& McWilliam 2008; Lee et al. 2008; Sbordone et al. 2010b). 
The cited precisions attainable are what we expect to obtain in
a survey conducted with such an instrument. 
 
\begin{itemize}
\item ${\rm T_{\rm eff}}$: stems from a combination of photometry, excitation equilibrium 
using \ion{Fe}{i} lines, and, in F--G dwarf stars, from the wings of the H$\alpha$ line. 
Temperatures can usually be settled to within 150\,K or better. 
\item \logg: employs ionisation equilibrium from \ion{Fe}{i} and \ion{Fe}{ii} and, if present,
other ionised and neutral species. The spectral range indicated above ensures the presence 
of several  ionised iron lines. 
Overall, a precision in the gravities within $\pm $0.5\,dex is realistic.
Ionisation equilibrium of other species can also be used, such as \ion{Ti}{i}/\ion{Ti}{ii}
or \ion{V}{i}/\ion{V}{ii}.
\item $\xi$: micro-turbulent velocities will be determined by removing any trend in 
abundance with line strength based on the large number of  \ion{Fe}{i} lines of various 
strength available in the spectra, to a precision of about 0.2\kms. 
\item {[Fe/H]}: more than 100 \ion{Fe}{i} lines are typically available in the envisioned 
target stars' spectra and our ample experience has shown that an accuracy of better than 0.1\,dex is easily achievable. 
\item {[$\alpha$/Fe]}: several Ca and Mg lines are measurable in the spectra 
(see Table\,\ref{range}) and precisions of 0.1\,dex are commonly reached. 
\item {[X/H]} for a variety of chemical species X: ideally, spectroscopic surveys foresee  
precisions reaching from 0.1\,dex (e.g., in [Ti/H]) to 0.5\,dex  (e.g., in [S/H], where feasible).
\end{itemize}

Given these constraints, we summarise the useful (i.e., unblended and/or recoverable) 
lines to be employed in a meaningful, high-resolution abundance analysis in Table\,\ref{range}.
Here, we list the number of lines that are potentially detectable in the
selected wavelength ranges, separately for dwarfs and giants, 
although we caution that not all lines will be detectable in all spectra due to their different sensitivities to the stellar parameters. 
Hence, in a typical survey strategy,  for those elements, for which only few lines are
available, one  cannot derive a chemical abundance ratio for the complete set of targeted stars.
The spectral content, summarised in Table\,\ref{range}, contains
diagnostic information for the determination of the atmospheric parameters.

In addition to the aforementioned Fe- and $\alpha$-element transitions, we  
will rely on the following absorption features 
for the abundance measurements of further, crucial tracers of Galactic chemical evolution: 
the G-band of CH (429--432\,nm), to derive the abundance of carbon; 
the CN band at 414--422\,nm will be used as an indicator of the N abundance once the C-abundance is known; 
the [OI] 630\,nm feature can serve to derive O-abundances in giants; 
the \ion{Na}{i}-lines at 589\,nm and  \ion{Al}{i} lines (at 396.152\,nm) cover representative $p$-capture elements; 
the \ion{Li}{i} doublet at 670.7\,nm is the only accessible means of determining A(Li). Finally, 
several transitions are available to  derive the abundance ratios for some $n$-capture elements.
Sr, Y, and Zr trace the first peak of $n$-capture elements and Ba, La, Ce, and Nd trace the second peak.
In the solar system abundances, all have been produced mainly through the slow $n$-capture process
($s$-process), although all have a rapid $n$-capture process ($r$-process) component that can be as large as
42\% in the case of Nd (Sneden et al. 2008). 
Finally also Eu, an almost pure $r$-process element, can be observed in the proposed 4MOST configuration.

We do not consider  in detail the G-band  in our further tests, 
since  this feature is clearly measurable also at low resolution, and thus would put no
stringent constraint on the design of the HR spectrograph.

We finally mention some interesting features that  one  looses with our trade-off choice of the  wavelength ranges.
\begin{itemize}
\item
The \ion{Mg}{i}-b triplet at about 516\,nm, which is useful for abundance determination in metal-poor stars,
$[{\rm M/H}]<-2.5$, and also for gravity determination for metal-rich stars. The large majority
of stars that 4MOST will observe will be close to solar metallicity, and,
at the resolving power of 4MOST,
the lines of the Mg\,b triplet are strongly saturated, rendering them not 
useful for abundance determination.
The wings of the saturated Mg\,b triplet lines would be a useful indicator for 
the stellar gravity, but our spectral range already contains a sufficient number of  gravity-sensitive features (Table\,\ref{range}) and
the majority of the stars, as stated above, will be selected
to be turn-off stars, so that their gravity will be known {\em a priori}.
\item
The oxygen triplet at 777\,nm is a very useful tool to derive the oxygen abundance,
 in dwarf stars at all metallicities. 
Oxygen abundance in giants will be derived from the [OI] line at 630\,nm.
For dwarf stars we will not be able to measure any O-abundance.
While including the  oxygen triplet would thus be highly valuable, the range around
it is very poor in lines so that no other information could be gleaned.
\item{}
The near-infrared \ion{Ca}{ii} triplet at about 850\,nm is included in the
range observed by Gaia.
We excluded these features from our HR wavelength ranges for the same reasons as for excluding 
the Mg\,b triplet and oxygen triplets, i.e.,  the lines are too strong to be
good abundance indicators in solar metallicity stars at the
resolving power of 4MOST and the range around
them has little other interesting lines.
Furthermore, that wavelength range  is highly contaminated by telluric absorption and night-sky emission lines.
\end{itemize}

%
\subsection{Spectral resolution and pixel-size}

In the design of a spectrograph, resolution and spectral coverage
are correlated. In order to avoid undersampling one needs at least two
pixels per resolution element. This implies that for
a detector of given physical dimension of the pixel and number of pixels 
increasing the resolution reduces the spectral coverage.
While a higher  resolution is desirable for a chemical abundance analysis,  
we also prefer a wavelength range that is as wide as possible. 
This should be coupled to an overall instrumental efficiency  
that allows for the detection of the spectral lines over
the spectral noise. 

Another aspect that influences the chemical line-by-line analysis
is pixel-size (in wavelength units, defined as pixel-size$=\lambda _{i+1}-\lambda _{i}$,
see Cayrel 1988): 
a small pixel-size allows one to better characterise the shape
of each line that is sampled by a large number of pixels. 
However, a larger number of pixels decreases the $S/N$ ratio per pixel,
because the same number of photons is spread over more pixels.
An unblended line, without hyperfine structure and/or isotopic
splitting, needs several pixels (from experience at least about five, 
one around the line core and two on each wing) for
its shape to be adequately described, in a way to allow line profile fitting
for the abundance analysis. 
The full line width depends on its intensity 
(see appendix A) so that weaker lines are sampled by a smaller number
of pixels. Thus
the smallest acceptable pixel-size is fixed by the weakest lines that
we want to measure.
For a fixed resolution
the spectral sampling (i.e., the number of  pixels
per  resolution element) 
changes with wavelength: \\
sampling = $\lambda$/(resolving power)/pixel-size.

It is self-evident that the instrumental characteristics must  be feasible from a technical point of view.
Therefore, we investigate in the following the cases of  a good compromise, at an average resolving power of 20\,000 and 
a minimum requirement of $R>$ 15\,000 across the entire spectral range, and a sampling between 2 and about 4,
which is sufficient to resolve the lines without dividing the photons between too many pixels.

\subsubsection{Weak versus strong lines}
As a first test, we investigated two cases: 
the strong \ion{Ba}{ii} line at 455.4\,nm ($6 {\rm pm} \le{\rm EW}\le 18$\,pm) and 
the \ion{Ti}{i} line at 454.8\,nm as an example of a weak line (EW=2.3\,pm 
for the metal-poor model with T$_{\rm eff}$/\logg/[M/H] = 5500/4.0/$-2.0$).

The \ion{Ba}{ii} line lies, especially in metal-rich stars,
in a crowded spectral region. 
Thus this feature  allows us to probe also
the effects of blending.
In practice, we fitted, minimising the $\chi ^2$, the line profile in a synthetic 
spectrum that has been injected with Poisson noise  
to mimic $S/N$ ratios of 50 and 40 per pixel at three different values for resolving powers 15\,000, 20\,000, and 25\,000 
and a pixel-size fixed at 2\,pm.  
The results are reported in Table\,\ref{baiitab}, where we list the recovered, mean abundance ratios and their standard deviations ($\sigma$).  
The latter were obtained by running 
1\,000 Monte Carlo trials. In some cases the fit was not successful, giving an abundance which deviates
from the input one by several orders of magnitudes, and we
reject these events, while keeping all others that we define as ``good events''.
As expected, as the resolution decreases, the numbers of times the fit fails increases,  as does the scatter $\sigma$. 
 
\begin{table*}[htb]                                                                                 
\begin{center}                                                                                      
\caption{Accuracy of the 455.4\,nm \ion{Ba}{ii} line as a function of resolution.
The pixel-size is 0.002\,nm. The $S/N$ is given per pixel.
}                               
\begin{tabular}{c c c r c}                                                                      
\hline                                                                                             
Model (T$_{\rm eff}$/\logg /[Fe/H] ) & Resolving Power & $S/N$  & Good events & [Ba/H]\\ \hline                              
\multicolumn{5}{c}{EW = 17.5 pm}              \\
\hline
4500/2.0/-1.0 & 15\,000 & 50  &  976/1000 & $-1.021\pm 0.065$ \\
4500/2.0/-1.0 & 20\,000 & 50  & 1000/1000 & $-1.027\pm 0.050$ \\
4500/2.0/-1.0 & 25\,000 & 50  & 1000/1000 & $-1.029\pm 0.047$ \\
\hline                                                
4500/2.0/-1.0 & 15\,000 & 40  &  972/1000 & $-1.017\pm 0.079$ \\
4500/2.0/-1.0 & 20\,000 & 40  & 1000/1000 & $-1.023\pm 0.061$ \\
4500/2.0/-1.0 & 25\,000 & 40  & 1000/1000 & $-1.026\pm 0.056$ \\
\hline 
\multicolumn{5}{c}{EW = 17.0 pm}              \\
\hline 
5500/4.0/+0.0 & 15\,000 & 50  &  996/1000 & $-0.013\pm 0.064$ \\
5500/4.0/+0.0 & 20\,000 & 50  &  999/1000 & $-0.012\pm 0.048$ \\
5500/4.0/+0.0 & 25\,000 & 50  & 1000/1000 & $-0.019\pm 0.039$ \\
\hline 
\multicolumn{5}{c}{EW = 11.6 pm}              \\
\hline 
5500/4.0/-1.0 & 15\,000 & 50  &  974/1000 & $-0.995\pm 0.071$ \\
5500/4.0/-1.0 & 20\,000 & 50  &  997/1000 & $-1.003\pm 0.051$ \\
5500/4.0/-1.0 & 25\,000 & 50  & 1000/1000 & $-0.997\pm 0.044$ \\
\hline 
\end{tabular} 
\label{baiitab} 
\end{center}
\end{table*}

In the next step, we investigated the effect that the combination of resolving power with a given pixel-size
has on the abundance determination.
For each of the above  resolving powers, we considered different samplings of 2.0, 2.5 and 3.0
pixels per resolution element. 
The resulting pixel-size was computed at a reference wavelength of 395\,nm.
In general, very strong lines
(${\rm EW}\ge 10$\,pm) are easily detectable above spectral noise even at faint magnitudes. 
However, such features are less suitable for a reliable  abundance determination, because they are difficult to model 
due to uncertain damping constants and usually stronger departures from Local Thermodynamic Equilibrium (LTE), 
and the growing importance of 3D effects (e.g., Caffau et al. 2011b). Furthermore, they are
significantly less sensitive to changes in abundance compared to weaker, unsaturated lines.
The EWs quoted in Table \ref{resps} refer to the suitably strong \ion{Ba}{ii} transition in the metal-poor model and the weaker \ion{Ti}{i} line in the same model spectrum.
 
\begin{table*}[htb]                                                                                   
\begin{center}                                                                                      
\caption{Accuracy of the 455.4\,nm \ion{Ba}{ii} line as a function of resolving power and pixel size. These tests are for a  strong line and the
454.8\,nm \ion{Ti}{i} line as a test case for a weak line. The $S/N$ is 160 per \AA~each.
}                               
\begin{tabular}{c c l c c c r c}
\hline                                                                                              
Line & Model & Pixel & Resolving Power & $S/N$  & Sampling & Good   & [X/H]\\ 
\relax
[nm] & (T$_{\rm eff}$/\logg /[Fe/H]) & Size [nm] &            & per pixel     &  @395\,nm & Events & \\
\hline\hline 
\multicolumn{8}{c}{EW = 6.6 pm}              \\
\hline                       
Ba II 455.4 & 5500/4.0/$-2.0$ & 0.0078  & 25\,000 & 45 & 2.0 &  977/1000 & $-1.965\pm 0.098$ \\
Ba II 455.4 & 5500/4.0/$-2.0$ & 0.00624 & 25\,000 & 40 & 2.5 &  976/1000 & $-1.969\pm 0.093$ \\
Ba II 455.4 & 5500/4.0/$-2.0$ & 0.0052  & 25\,000 & 36 & 3.0 &  985/1000 & $-1.973\pm 0.091$ \\
   \hline                                                        
Ba II 455.4 & 5500/4.0/$-2.0$ & 0.00975 & 20\,000 & 50 & 2.0 &  979/1000 & $-1.959\pm 0.113$ \\
Ba II 455.4 & 5500/4.0/$-2.0$ & 0.0078  & 20\,000 & 45 & 2.5 &  971/1000 & $-1.964\pm 0.115$ \\
Ba II 455.4 & 5500/4.0/$-2.0$ & 0.0065  & 20\,000 & 41 & 3.0 &  973/1000 & $-1.970\pm 0.111$ \\
   \hline                                                        
Ba II 455.4 & 5500/4.0/$-2.0$ & 0.0130  & 15\,000 & 58 & 2.0 &  929/1000 & $-1.950\pm 0.157$ \\
Ba II 455.4 & 5500/4.0/$-2.0$ & 0.0104  & 15\,000 & 52 & 2.5 &  938/1000 & $-1.950\pm 0.162$ \\
Ba II 455.4 & 5500/4.0/$-2.0$ & 0.0087  & 15\,000 & 47 & 3.0 &  903/1000 & $-1.953\pm 0.162$ \\
   \hline                                                       
\multicolumn{8}{c}{EW = 2.0 pm}              \\
\hline                       
Ti I 454.8  & 5500/4.0/$-2.0$ & 0.0078  & 25\,000 & 45 & 2.0 &  814/1000 & $-2.002\pm 0.137$ \\
Ti I 454.8  & 5500/4.0/$-2.0$ & 0.00624 & 25\,000 & 40 & 2.5 &  763/1000 & $-2.000\pm 0.150$ \\
Ti I 454.8  & 5500/4.0/$-2.0$ & 0.0052  & 25\,000 & 36 & 3.0 &  836/1000 & $-2.011\pm 0.142$ \\
   \hline                                                         
Ti I 454.8  & 5500/4.0/$-2.0$ & 0.00975 & 20\,000 & 50 & 2.0 &  741/1000 & $-2.154\pm 0.237$ \\
Ti I 454.8  & 5500/4.0/$-2.0$ & 0.0078  & 20\,000 & 45 & 2.5 &  764/1000 & $-2.008\pm 0.180$ \\
Ti I 454.8  & 5500/4.0/$-2.0$ & 0.0065  & 20\,000 & 41 & 3.0 &  640/1000 & $-2.027\pm 0.202$ \\
   \hline                                                         
Ti I 454.8  & 5500/4.0/$-2.0$ & 0.0130  & 15\,000 & 58 & 2.0 &  645/1000 & $-1.973\pm 0.301$ \\
Ti I 454.8  & 5500/4.0/$-2.0$ & 0.0104  & 15\,000 & 52 & 2.5 &  651/1000 & $-1.973\pm 0.258$ \\
Ti I 454.8  & 5500/4.0/$-2.0$ & 0.0087  & 15\,000 & 47 & 3.0 &  597/1000 & $-1.979\pm 0.315$ \\
\hline 
\end{tabular} 
\label{resps} 
\end{center}
\end{table*} 

As these tests indicate, a strong line is much less affected by a change in resolving power than a weak line.
In contrast,  changes in pixel-size at a given  $S/N$ ratio (stated per \AA) 
have no significant effect on the abundance determination.
A consequence of fixing the  $S/N$ ratio per \AA, a decrease in pixel-size leads 
to a better sampling  of the line profile, with a higher number of pixels that
resolve the line, but at the expense of a lower $S/N$ ratio in each pixel.
In conclusion, we suggest as a compromise in 
 resolving power and pixel sampling one of the following combinations: 
20\,000/2.5,  25\,000/2.0--2.5, or 15\,000/2.5--3.0.

\subsubsection{Particular lines}
In the following we briefly reflect upon a few elements of particular importance (Sect.~2.4), for which abundances need to be derived 
from essentially one line and which require special care. 

Oxygen is a fundamental tracer of fast chemical enrichment that can efficiently 
be derived in giants, but poses several problems in dwarf spectra.
The forbidden  [OI] line at 630\,nm (see Table\,\ref{oitab}), available in our wavelength range
(and a typical EW of about 0.35\,pm in the solar spectrum; Caffau et al. 2008) is too weak in dwarf stars.
The line becomes stronger in giants and can be measured down to [Fe/H]$< -3$ (e.g., Cayrel et al. 2004). 
Some telluric absorptions affect the region, and, for stars with low radial velocity,
the atmospheric emission can also affect the line.
Even if the line is problematic, it is very important to derive oxygen abundance,
at least for a fraction of stars.
As our tests show, a resolving power of at least 20\,000 is necessary to derive 
the oxygen abundance from this [OI] line, see Table\,\ref{oitab}.
In fact,  our simulations indicate that a resolving power of 15\,000 
yields only  60\% ``good events'' 
(for the case of the model at [M/H]=$-$1.0), compared to  
a success rate of  90\% at a resolving power of 25\,000.

\begin{table*}[htb]
\begin{center}                                                                                     
\caption{The same as Table\,\ref{resps}, but for  the [OI] line at 630.0\,nm.}                               
\begin{tabular}{c c l c c c r c}
\hline                                                                                           
Line & Model & Pixel & Resolving Power & $S/N$  & Sampling & Good   & [O/H]\\ 
\relax
[nm] & (T$_{\rm eff}$/\logg /[Fe/H]) & Size [nm] &            &  per pixel    &  @395\,nm&Events & \\
\hline                       
\multicolumn{8}{c}{EW = 4.8 pm}              \\
\hline                      
$\left[{\rm OI}\right]$ 630.0 & 4500/2.0/0.0 & 0.0078  & 25\,000 & 45 & 2.0 &  919/1000 & $ 0.013\pm 0.104$ \\
$\left[{\rm OI}\right]$ 630.0 & 4500/2.0/0.0 & 0.00624 & 25\,000 & 40 & 2.5 &  920/1000 & $ 0.019\pm 0.108$ \\
$\left[{\rm OI}\right]$ 630.0 & 4500/2.0/0.0 & 0.0052  & 25\,000 & 36 & 3.0 &  942/1000 & $ 0.016\pm 0.108$ \\
\hline                                                       
$\left[{\rm OI}\right]$ 630.0 & 4500/2.0/0.0 & 0.00975 & 20\,000 & 50 & 2.0 &  905/1000 & $ 0.015\pm 0.144$ \\
$\left[{\rm OI}\right]$ 630.0 & 4500/2.0/0.0 & 0.0078  & 20\,000 & 45 & 2.5 &  904/1000 & $ 0.020\pm 0.133$ \\
$\left[{\rm OI}\right]$ 630.0 & 4500/2.0/0.0 & 0.0065  & 20\,000 & 41 & 3.0 &  920/1000 & $ 0.026\pm 0.133$ \\
\hline                                      
$\left[{\rm OI}\right]$ 630.0 & 4500/2.0/0.0 & 0.0130  & 15\,000 & 58 & 2.0 &  893/1000 & $ 0.036\pm 0.198$ \\
$\left[{\rm OI}\right]$ 630.0 & 4500/2.0/0.0 & 0.0104  & 15\,000 & 52 & 2.5 &  892/1000 & $ 0.042\pm 0.182$ \\
$\left[{\rm OI}\right]$ 630.0 & 4500/2.0/0.0 & 0.0087  & 15\,000 & 47 & 3.0 &  882/1000 & $ 0.038\pm 0.201$ \\
\hline                  
\multicolumn{8}{c}{EW = 0.8 pm}              \\
\hline                
$\left[{\rm OI}\right]$ 630.0 & 4500/2.0/-1.0 & 0.0078  & 25\,000 & 45 & 2.0 &  792/1000 & $-0.974\pm 0.185$ \\
$\left[{\rm OI}\right]$ 630.0 & 4500/2.0/-1.0 & 0.00624 & 25\,000 & 40 & 2.5 &  845/1000 & $-0.973\pm 0.188$ \\
$\left[{\rm OI}\right]$ 630.0 & 4500/2.0/-1.0 & 0.0052  & 25\,000 & 36 & 3.0 &  838/1000 & $-0.964\pm 0.180$ \\
\hline                                                  
$\left[{\rm OI}\right]$ 630.0 & 4500/2.0/-1.0 & 0.00975 & 20\,000 & 50 & 2.0 &  749/1000 & $-0.941\pm 0.220$ \\
$\left[{\rm OI}\right]$ 630.0 & 4500/2.0/-1.0 & 0.0078  & 20\,000 & 45 & 2.5 &  695/1000 & $-0.967\pm 0.231$ \\
$\left[{\rm OI}\right]$ 630.0 & 4500/2.0/-1.0 & 0.0065  & 20\,000 & 41 & 3.0 &  759/1000 & $-0.940\pm 0.215$ \\
\hline                                      
$\left[{\rm OI}\right]$ 630.0 & 4500/2.0/-1.0 & 0.0130  & 15\,000 & 58 & 2.0 &  609/1000 & $-0.790\pm 0.329$ \\
$\left[{\rm OI}\right]$ 630.0 & 4500/2.0/-1.0 & 0.0104  & 15\,000 & 52 & 2.5 &  623/1000 & $-0.778\pm 0.349$ \\
$\left[{\rm OI}\right]$ 630.0 & 4500/2.0/-1.0 & 0.0087  & 15\,000 & 47 & 3.0 &  590/1000 & $-0.772\pm 0.339$ \\
\hline 
\hline
\end{tabular} 
\label{oitab}
\end{center}
\end{table*}

The lithium abundance is mostly derived from the 670.7\,nm doublet.
Our experience has shown that this feature can be detected when the spectral resolving power
is 20\,000, provided a $S/N$ ratio of about 80 per pixel 
\cite[see e.g.][]{M10,M12}. 
With the spectrograph design we are testing in this work, this S/N ratio is achieved only
for the brightest objects.
Nevertheless, this will be sufficient to derive Li abundance for a large sample of Galactic
stars to allow us to improve our understanding on the history of lithium in the MW.

Strontium is the first of the weak s-process neutron-capture
elements at Solar metallicity (Heil et al. 2009, Pignatari et al. 2008, and Arlandini et al. 1999),
which we can detect. Due to the strong resonance lines (in particular the
407.7\,nm line) the detection of this element sets no constraint on the
required resolution. It remains measurable in both dwarfs and giants at
all metallicities. However, in the stars 
[Fe/H] $> -2.0$
the measurement of this line poses a problem in an automatic fitting
routine owing to the strong line blends in both Sr wings, and the fact that
the line saturates at even higher metallicities.

As an example for a rapid $n$-capture element, we considered 
Eu and ran our tests on the \ion{Eu}{ii} line at 
664.5\,nm\,(see Tables\,\ref{range} and \ref{euiitab}).
Europium is a second peak neutron-capture element. 
It is mainly synthesised by $r$-process in massive stars
(not yet fully understood -- for details see Fran\c cois et al. 2007).
As our tests indicate, this procedure yields reliable results both for dwarf and giant stars so that no stringent 
limitations in resolution is required in this case. 

\begin{table*}[htb]                                                                           
\begin{center}                                                                                      
\caption{The same as Table\,\ref{oitab}, but for  the Eu II  line at 664.5\,nm.}                               
\begin{tabular}{ c c l c c c r c }
\hline                                                                   
Line & Model & Pixel & Resolving Power & $S/N$  & Sampling & Good   & [Eu/H]\\ 
\relax
[nm] & (T$_{\rm eff}$/\logg /[Fe/H]) & Size [nm] &            &  per pixel     &  @395\,nm&Events & \\
\hline     
\multicolumn{8}{c}{EW = 5.0 pm}              \\
\hline    
Eu II 664.5 & 4500/2.0/0.0 & 0.0078  & 25\,000 & 45 & 2.0 &  901/1000 & $ 0.000\pm 0.101$ \\
Eu II 664.5 & 4500/2.0/0.0 & 0.00624 & 25\,000 & 40 & 2.5 &  896/1000 & $ 0.001\pm 0.101$ \\
Eu II 664.5 & 4500/2.0/0.0 & 0.0052  & 25\,000 & 36 & 3.0 &  913/1000 & $ 0.002\pm 0.108$ \\
\hline                                                       
Eu II 664.5 & 4500/2.0/0.0 & 0.00975 & 20\,000 & 50 & 2.0 &  915/1000 & $ 0.001\pm 0.116$ \\
Eu II 664.5 & 4500/2.0/0.0 & 0.0078  & 20\,000 & 45 & 2.5 &  927/1000 & $ 0.001\pm 0.118$ \\
Eu II 664.5 & 4500/2.0/0.0 & 0.0065  & 20\,000 & 41 & 3.0 &  906/1000 & $-0.001\pm 0.111$ \\
\hline                                           
Eu II 664.5 & 4500/2.0/0.0 & 0.0130  & 15\,000 & 58 & 2.0 &  836/1000 & $ 0.028\pm 0.136$ \\
Eu II 664.5 & 4500/2.0/0.0 & 0.0104  & 15\,000 & 52 & 2.5 &  892/1000 & $ 0.014\pm 0.136$ \\
Eu II 664.5 & 4500/2.0/0.0 & 0.0087  & 15\,000 & 47 & 3.0 &  882/1000 & $ 0.014\pm 0.139$ \\
\hline                       
\multicolumn{8}{c}{EW = 10.0 pm}              \\
\hline                      
Eu II 412.9 & 4500/2.0/$-2.0$ & 0.0078  & 25\,000 & 45 & 2.0 &  952/1000 & $-1.941\pm 0.394$ \\
Eu II 412.9 & 4500/2.0/$-2.0$ & 0.00624 & 25\,000 & 40 & 2.5 &  945/1000 & $-1.951\pm 0.337$ \\
Eu II 412.9 & 4500/2.0/$-2.0$ & 0.0052  & 25\,000 & 36 & 3.0 &  969/1000 & $-1.961\pm 0.302$ \\
\hline                                                         
Eu II 412.9 & 4500/2.0/$-2.0$ & 0.00975 & 20\,000 & 50 & 2.0 &  900/1000 & $-1.954\pm 0.273$ \\
Eu II 412.9 & 4500/2.0/$-2.0$ & 0.0078  & 20\,000 & 45 & 2.5 &  953/1000 & $-1.956\pm 0.311$ \\
Eu II 412.9 & 4500/2.0/$-2.0$ & 0.0065  & 20\,000 & 41 & 3.0 &  958/1000 & $-1.942\pm 0.365$ \\
\hline                                             
Eu II 412.9 & 4500/2.0/$-2.0$ & 0.0130  & 15\,000 & 58 & 2.0 &  823/1000 & $-1.893\pm 0.554$ \\
Eu II 412.9 & 4500/2.0/$-2.0$ & 0.0104  & 15\,000 & 52 & 2.5 &  931/1000 & $-1.936\pm 0.393$ \\
Eu II 412.9 & 4500/2.0/$-2.0$ & 0.0087  & 15\,000 & 47 & 3.0 &  933/1000 & $-1.936\pm 0.400$ \\
\hline 
\hline 
\end{tabular} 
\label{euiitab} 
\end{center}
\end{table*}
 
\subsection{Precision in chemical abundances}
We would like to stress that the precision we are here considering
is only referring to the random errors based on the noise characteristics of the spectra. 
No systematic error related to uncertainties on  stellar parameters (T$_{\rm eff}$, \logg, $\xi$, [M/H]) 
and atomic data ($gf$-values) will be  taken into account, neither other noise sources.

With millions of individual stars, the amount of spectral data that present-day surveys will observe is too large for manual
 abundance analysis \citep[e.g.][]{cayrel04,bonifacio09}.   
Thus, an automatic approach for deriving the stellar parameters and the chemical compositions
of the observed stars is necessary. 
Only a few species (e.g.,  Li, S, Sr, and Th) require an {\em ad hoc} analysis approach. For instance, Li 
requires a NLTE treatment, Sr is heavily blended, and the Th line at 408.6\,nm 
lies in the red wing of a stronger blend that has to be taken into account properly.

\begin{figure*}[htb]
   \includegraphics[width=13cm]{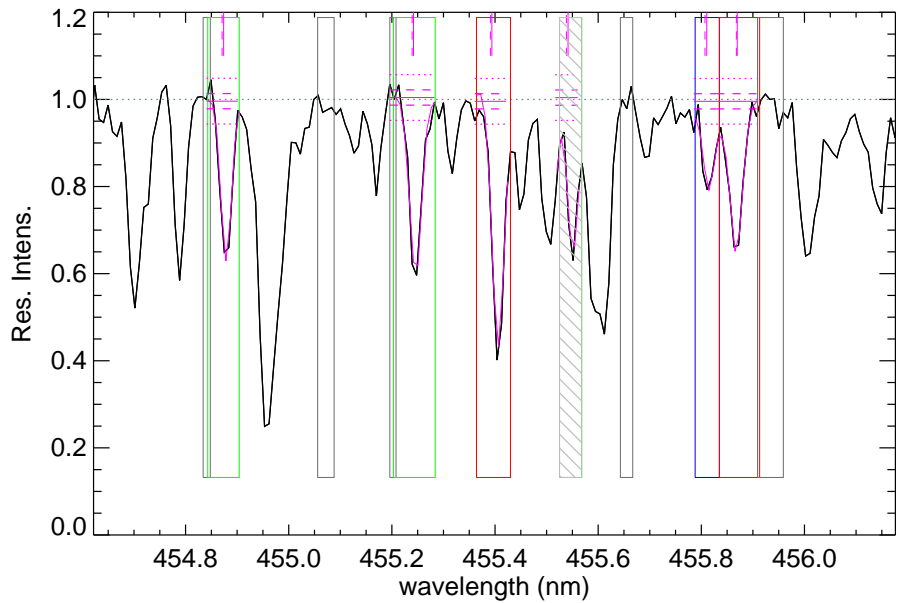} 
 \caption{A portion of the spectrum of a turn-off star with Solar-scaled abundances.
 Gray boxes show the pseudo-continuum window. 
 while coloured boxes mark windows around several absorption 
 features: \ion{Fe}{i} (blue), $\alpha$-elements (green), and heavy elements (red).
 Here, the red boxes show the fits of the \ion{Ba}{ii} line at 455.4\,nm
 and of the \ion{Cr}{ii} feature at 455.8\,nm. The gray shaded line is rejected because the
 $\chi ^2$ is bellow the fixed threshold.
 Magenta horizontal lines: solid lines represent the determined continuum,
 dashed lines represent one $\sigma$ of the S/N ratio,
 dotted lines represent three $\sigma$ of the S/N ratio.
 Magenta vertical lines show the shift performed by the fitting procedure.}
    \label{fig:mygi}
\end{figure*}

In the present analysis, we used the code MyGIsFOS (Sbordone et al. 2010a) to derive the chemical composition of
our test cases.
MyGIsFOS performs a fully  automated abundance analysis.
From a number of predefined regions, MyGIsFOS performs a ``pseudo-normalisation''  
of both the input spectrum to be analysed and the grid of synthetic spectra. 
The chemical abundance analysis is then performed by line profile fitting of a sample of selected lines. 
In Fig.\,\ref{fig:mygi}, an example of the analysis of MyGIsFOS is shown.
One of the advantages of the line-profile-fitting technique is that, 
in principle, one is free from uncertainties linked to blends.
This is true in the present analysis since the atomic data of the simulations
are exactly the same as the ones of the synthetic grid.
In reality, the blends introduce an uncertainty due to poor knowledge 
of the atomic data of the blending lines and the potential  presence of unknown blends.
An assessment of this kind of uncertainty is beyond the scope of the present analysis.

\begin{figure}[htb]
   \includegraphics[width=8cm,clip=true]{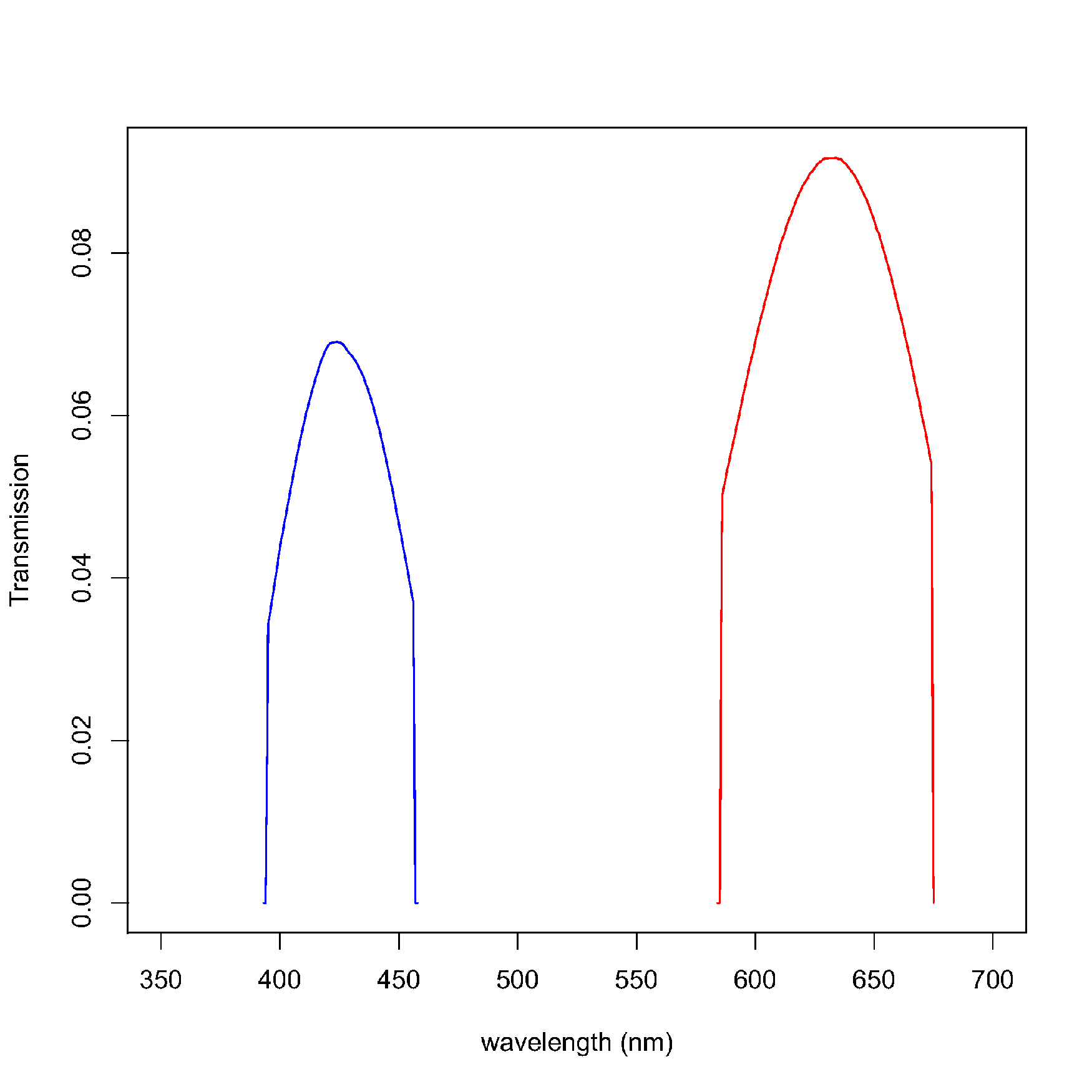} 
 \caption{The transmission of the instrument plus the telescope.}
    \label{qe}
\end{figure}

With MyGIsFOS we analysed the spectra obtained from the 4MOST simulator,
computed with the characteristics given in Table\,\ref{design} (the overall
transmission of the instrument plus telescope is shown in Fig.\ref{qe})
to derive the detailed chemical abundances, while temperature, gravity, and micro-turbulence
are kept fixed at the value of the input simulated synthetic spectrum.

As an example of our procedure, we report in Tables\,\ref{tab:dmr} and \ref{tab:gmp} 
the chemical abundances we derive 
for a Solar metallicity turn-off star (5500/4.0/0.0) and a metal-poor giant
(4500/2.0/$-$3.0), both at an adopted magnitude of  $r=14$\,mag and simulated to be observed for 3\,200\,s,
which is a representative exposure time in 4\,m class telescope survey strategies. 

The turn-off star of Solar metallicity is a typical star of the Galactic disc.
The line-to-line scatters\footnote{That is, the standard deviation of the abundance measurements from all lines of one element}, 
$\sigma$, in Table\,\ref{tab:dmr} are mostly within 0.1\,dex for all elements, except for Zr,
for which the abundance determination would greatly benefit 
from an  increase in exposure time.
 
\begin{table}[htb]                                                                                  
\begin{center}                                                                                     
\caption{Chemical abundance results for a solar metallicity turn-off star (5500\,K/4.0/0.0), with $r=14$\,mag.
$\sigma$ denotes the line-to-line scatter.}   
\begin{tabular}{l l l l }                                
\hline                                                       
 Element & N lines & [X/H] & $\sigma$\\
\hline                                                
 \ion{Na}{i} &  2 & --0.10 & 0.02\\
 \ion{Mg}{i} &  2 & --0.01 & 0.06\\
 \ion{Al}{i} &  2 & --0.13 & 0.03\\
 \ion{Si}{i} & 11 & --0.06 & 0.04\\
 \ion{Ca}{i} & 16 & --0.03 & 0.04\\
 \ion{Sc}{i}/\ion{Sc}{ii} &  2/4 & +0.07/--0.10 & 0.11/0.07\\
 \ion{Ti}{i}/\ion{Ti}{ii} & 18/6 & --0.04/--0.04 & 0.07/0.06\\
 \ion{V}{i}/\ion{V}{ii} & 13/1 & --0.04/--0.25 & 0.07\\
 \ion{Cr}{i}/\ion{Cr}{ii} & 3/1 & +0.05/--0.011 & 0.03\\
 \ion{Mn}{i}  & 13 & --0.11 & 0.03\\
 \ion{Fe}{i}/\ion{Fe}{i} & 116/11 & --0.07/--0.12 & 0.06/0.09\\
 \ion{Co}{i}  &  6 & --0.07 & 0.07\\
 \ion{Ni}{i}  & 16 & --0.05 & 0.05\\
 \ion{Y}{ii}  &  2 & --0.10 & 0.02\\
 \ion{Zr}{ii} &  2 &  +0.14 & 0.35\\
 \ion{Ba}{ii} &  2 & --0.13 & 0.02\\
 \ion{La}{ii} &  1 & --0.08 & \dots \\
 \ion{Nd}{ii} &  2 & --0.08 & 0.07 \\
 \ion{Eu}{ii} &  1 & --0.14 & \dots \\
\hline                                                
\end{tabular} 
\label{tab:dmr} 
\end{center}
\end{table}

For the Galactic halo star (Table\,\ref{tab:gmp}) the considerations are necessarily different.
The average metallicity of inner halo stars is $-1.6$\,dex (Carollo et al. 2007).
When observing giants,
one can observe stars further away from the Sun and thereby probe the halo
star composition.
One of the goals in the target selection of halo stars is to detect and characterise the most
metal-poor stars in the Galactic halo. 
The lower the content of metals in a star the weaker the metallic lines appear in the spectrum.
Thus we decided to test as an extreme case the most metal-poor giant,  for which we computed a synthetic
spectrum with $\left[{\rm M/H}\right]=-3.0$. The results will then represent a lower limit to 
the precision compared to what can be achieved for stars 10--100 times more metal-rich. 
 
\begin{table}[htb]                                                                                   
\begin{center}                                                                                     
\caption{The same as Table\,\ref{tab:dmr}, but for a  giant star (4500\,K/2.0/$-$3.0), with $r=14$\,mag.}   
\begin{tabular}{ l l l l  }                                
\hline
 Element & N lines & [X/H] & $\sigma$\\
\hline                                                
 \ion{Na}{i}  &  2 & --3.13 & 0.03\\
 \ion{Mg}{i}  &  1 & --2.55 & \dots \\
 \ion{Al}{i}  &  1 & --2.99 & \dots \\
 \ion{Si}{i}  &  2 & --2.42 & 0.20\\
 \ion{Ca}{i}  & 13 & --2.60 & 0.04\\
 \ion{Sc}{ii} &  4 & --3.11 & 0.20\\
 \ion{Ti}{i}/\ion{Ti}{ii} & 15/19 & --2.58/--2.57 & 0.08/0.07\\
 \ion{V}{i}  & 3 &--2.96 & 0.02\\
 \ion{Cr}{i} & 3 &--2.97 & 0.05\\
 \ion{Mn}{i}/\ion{Mn}{ii} & 7/1  & --3.00/--2.93 & 0.05\\
 \ion{Fe}{i}/\ion{Fe}{ii} & 63/4 & --2.99/--2.84 & 0.07/0.11\\
 \ion{Co}{i}  & 4 & --3.06 & 0.07\\
 \ion{Ni}{i}  & 4 & --2.88 & 0.10\\
 \ion{Sr}{ii} & 1 & --3.08 &  \dots\\
 \ion{Y}{ii}  & 1 & --3.08 &  \dots\\
 \ion{Zr}{ii} & 2 & --3.01 &  \dots \\
 \ion{Ba}{ii} & 2 & --3.02 & 0.07\\
 \ion{La}{ii} & 5 & --2.92 & 0.23\\
 \ion{Nd}{ii} & 1 & --2.93 &  \dots\\
 \ion{Eu}{ii} & 1 & --3.11 &   \dots \\                         
\hline                                                
\end{tabular} 
\label{tab:gmp}
\end{center}
\end{table} 

The line-to-line scatter listed in Tables\,\ref{tab:dmr} and \ref{tab:gmp},  
both for the solar turn-off and for the metal-poor giant stars, is within 0.1\,dex  for the large majority of the elements.
It becomes significantly larger than 0.1\,dex for Zr in the turn-off spectrum,
where there are only two lines of \ion{Zr}{ii}. The same holds for 
Si, Sc, and La for the giant. 
For some elements a systematic effect is visible, in that the average 
abundance derived from the analysis differs from the known input value.
This effect is more evident for elements whose abundance is based on a single line.

A similar situation holds for the faintest targets at $r=16$\,mag, assuming
an exposure time of 6800\,s. This  results in a small increase of the uncertainties. 
We analysed also a metal-poor turn-off star (5500/4.0/$-2.0$),
that shows in its spectrum lines weaker than the same lines in the solar-metallicity spectrum.
As expected, the uncertainties we derive (see Table\,\ref{tab:dmr})  
are larger than in the solar-metallicity case for some elements

%
%
%
\subsection{Monte Carlo simulation on the complete wavelength range}
In the previous sections, we analysed single simulated spectra of individual targets, computed by the 4MOST
simulator. 
A single case of noise-injection can be useful to obtain a first, overall characterisation
of the performance of a spectrograph under a fixed design, e.g.
the minimum EW measurable in a line as a function of the S/N ratio, but it
is generally not sufficient to obtain a careful 
understanding on the final precision achievable, especially for those elements 
where abundances are based on a few lines only (Table\,\ref{range}).

Therefore, with the 4MOST simulator we computed a series of 1\,000 simulations of the same spectrum,
based on the values reported in Table\,\ref{design}. These were analysed with a Monte Carlo technique and employing the  
full range in wavelength to get an estimate of the  overall 
performance of a high resolution spectroscopic survey.
In real astronomical observations it is customary to estimate
the precision on the abundance of a given element from the line-to-line scatter.
The Monte Carlo simulation provides the mean line-to-line scatter and its dispersion,
over the sample, for all elements whose abundance can be derived from several lines.
For all elements the dispersion on the abundance, over the sample is also available.
We consider the former, when about ten lines of the element 
are measurable, as an estimate of the achievable precision,
resorting to the latter for elements measurable from a few lines. 
The two estimates are clearly connected, the latter being roughly
the former divided by the square root of the number of lines.

The results are shown in Table\,\ref{tab:mc_dmr}, where we report the results for a solar metallicity
turn-off star, and in Table\,\ref{tab:mc_dmp} where we show the recovered abundances for a metal-poor
turn-off star.
In the latter case, Cr, Eu, Nd, and ionised Si, present in the solar metallicity case, cannot be analysed, since the lines become too weak.
Vice versa, Sr is not measurable in an automated way in the solar metallicity star,  
due to the excessive blending, while it can be  measured automatically in the low metallicity
case.
Generally, the average line-to-line scatter is about a factor of two larger for the metal-poor case,
but there are exceptions, such as Ba, whose line-to-line scatter is smaller in the metal-poor model, 
or \ion{Fe}{ii}, for which both models show the same value.

\begin{table}[htb]     
\begin{center}       
\caption{Abundances results and uncertainties derived from 1\,000 Monte-Carlo simulations for the (5500/4.0/0.0) model
with $r=14$\,mag.
In the first column the elements analysed are listed. The following three columns refers to 1\,000 Monte Carlo simulation
events with an exposure time of 3\,200\,s, and are the minimum-maximum number of lines used for the abundance determination,
the average abundance over the 1\,000 Monte Carlo events, with its scatter, and the average line-to-line scatter
over the 1\,000 events with related standard deviation.}   
\begin{tabular}{l c r c}                                
\hline
 Element & N lines & $\langle\left[{\rm X/H}\right]\rangle$ & $\sigma$\\
\hline                                                
\hline                                                
 \ion{Na}{i}  & 2         & $-0.042\pm 0.058$ & $0.026\pm 0.020$\\
 \ion{Mg}{i}  & 1--2     & $ 0.026\pm 0.035$ & $0.042\pm 0.031$\\
 \ion{Al}{i}  & 1--2     & $ 0.038\pm 0.055$ & $0.050\pm 0.038$\\
 \ion{Si}{i}  & 6--14    & $-0.038\pm 0.027$ & $0.054\pm 0.014$\\
 \ion{Si}{ii} & 1--2     & $ 0.014\pm 0.074$ & $0.083\pm 0.064$\\
 \ion{Ca}{i}  & 8--16    & $-0.017\pm 0.022$ & $0.045\pm 0.009$\\
 \ion{Sc}{ii} & 2--5     & $-0.010\pm 0.055$ & $0.080\pm 0.033$\\
 \ion{Ti}{i}  & 8--18    & $-0.036\pm 0.026$ & $0.060\pm 0.014$\\
 \ion{Ti}{ii} & 2--7     & $ 0.018\pm 0.041$ & $0.057\pm 0.028$\\
 \ion{V}{i}   & 7--14    & $-0.043\pm 0.021$ & $0.062\pm 0.014$\\
 \ion{Cr}{i}  & 2--4     & $-0.034\pm 0.048$ & $0.042\pm 0.026$\\
 \ion{Cr}{ii} & 0--1     & $ 0.028\pm 0.108$ & \dots \\
 \ion{Mn}{i}  & 3--12    & $-0.064\pm 0.039$ & $0.061\pm 0.021$\\
 \ion{Fe}{i}  & 98--128  & $-0.056\pm 0.009$ & $0.070\pm 0.005$\\
 \ion{Fe}{ii} & 7--16    & $-0.050\pm 0.036$ & $0.101\pm 0.020$\\
 \ion{Co}{i}  & 2--6     & $-0.089\pm 0.037$ & $0.039\pm 0.028$\\
 \ion{Ni}{i}  & 8--20    & $-0.056\pm 0.024$ & $0.051\pm 0.011$\\
 \ion{Y}{ii}  & 1--2     & $-0.031\pm 0.068$ & $0.048\pm 0.046$\\
 \ion{Zr}{ii} & 1--2     & $-0.013\pm 0.067$ & $0.071\pm 0.065$\\
 \ion{Ba}{ii} & 0--2     & $ 0.014\pm 0.085$ & $0.086\pm 0.068$\\
 \ion{La}{ii} & 0--1     & $-0.082\pm 0.111$ &  \dots \\
 \ion{Nd}{ii} & 1--2     & $-0.036\pm 0.091$ & $0.080\pm 0.078$\\
 \ion{Eu}{ii} & 0--1     & $-0.023\pm 0.048$ & \dots  \\
\hline                                                
\end{tabular} 
\label{tab:mc_dmr}
\end{center}
\end{table} 

\begin{table}[htb]     
\begin{center}       
\caption{Abundances results and uncertainties derived from 1\,000 Monte-Carlo simulations 
for the (5500/4.0/$-2.0$) model and $r=14$\,mag. 
The exposure time, calculation method, and the meaning of the columns are the same
as in Table\,\ref{tab:mc_dmr}.}   
\begin{tabular}{l c r c}                                
\hline
 Element & N lines & $\langle\left[{\rm X/H}\right]\rangle$ & $\sigma$\\
\hline                                                
\hline                                                
 \ion{Na}{i}  & 0--2     & $-1.856\pm 0.565$ & $0.604\pm 0.425$\\
 \ion{Mg}{i}  & 0--2     & $-1.645\pm 0.077$ & $0.077\pm 0.062$\\ 
 \ion{Si}{i}  & 2--8     & $-1.561\pm 0.078$ & $0.105\pm 0.091$\\
 \ion{Ca}{i}  & 8--18    & $-1.619\pm 0.032$ & $0.069\pm 0.016$\\
 \ion{Sc}{ii} & 1--4     & $-1.974\pm 0.302$ & $0.247\pm 0.328$\\
 \ion{Ti}{i}  & 2--11    & $-1.561\pm 0.062$ & $0.108\pm 0.049$\\
 \ion{Ti}{ii} & 2--7     & $-1.593\pm 0.080$ & $0.081\pm 0.041$\\
 \ion{V}{i}   & 1--5     & $-1.985\pm 0.126$ & $0.101\pm 0.132$\\
 \ion{Mn}{i}  & 2--12    & $-2.056\pm 0.063$ & $0.107\pm 0.038$\\
 \ion{Fe}{i}  &  40--71  & $-2.006\pm 0.021$ & $0.135\pm 0.027$\\
 \ion{Fe}{ii} & 2--8     & $-2.006\pm 0.078$ & $0.102\pm 0.052$\\
 \ion{Co}{i}  & 1--7     & $-2.068\pm 0.064$ & $0.057\pm 0.031$\\
 \ion{Ni}{i}  &  0--5    & $-1.814\pm 0.235$ & $0.175\pm 0.229$\\
 \ion{Sr}{ii} & 0--1      & $-2.207\pm 0.118$ &  \dots \\
 \ion{Y}{ii}  & 1--2     & $-2.062\pm 0.082$ & $0.047\pm 0.045$\\
 \ion{Zr}{ii} & 0--1     & $-1.439\pm 0.081$ & \dots \\
 \ion{Ba}{ii} & 1--2     & $-2.038\pm 0.111$ & $ 0.057\pm 0.050$\\
 \ion{La}{ii} & 0--1     & $-1.522\pm 0.118$ &  \dots \\
\hline                                                
\end{tabular} 
\label{tab:mc_dmp}
\end{center}
\end{table} 

As a result,  in the case of the Solar turn-off star, 
we recover the abundance ratios of the majority of the elements 
systematically lower by typically a few hundredths of dex 
compared to the input spectrum of known parameters and input abundances. 
We also ran the Monte Carlo tests for the metal-poor giant with two different values of exposure time, 3\,200 and 6\,000\,s 
(see Table\,\ref{tab:mc_gmp}),  and for
the solar-metallicity giant (see Table\,\ref{tab:mc_gmr}).
As expected, by increasing the integration time, the scatter of the average value decreases, as well
as the average line-to-line scatter.
Twenty elements can be detected in the halo giant, and for five elements we can derive the ionisation equilibrium.

\begin{table*}[htb]     
\begin{center}       
\caption{Abundances results and uncertainties derived from 1\,000 Monte-Carlo simulations for the 
(4500/2.0/$-3.0$) model and $r=14$\,mag. 
Columns are as in Table\,\ref{tab:mc_dmr}, but there are two Monte Carlo simulations for two value of exposure time
of 3\,200\,s and 6\,000\,s, respectively.}
\begin{tabular}{l | c r c | c r c }                                
\hline
 Element & N lines & $\langle\left[{\rm X/H}\right]\rangle$ & $\sigma$ & N lines & $\langle\left[{\rm X/H}\right]\rangle$ & $\sigma$ \\
                   & \multicolumn{3}{c}{Exposure time of 3\,200 s}       & \multicolumn{3}{c}{Exposure time of 6\,000 s}                 \\
\hline                                                
\hline                                                
 \ion{Na}{i}  & 0--1     & $-3.170\pm 0.054$ &  \dots           & 0--1     & $-3.178\pm 0.044$ &  \dots          \\
 \ion{Mg}{i}  & 0--1     & $-2.527\pm 0.110$ &  \dots           & 0--1     & $-2.550\pm 0.074$ &  \dots          \\ 
 \ion{Si}{i}  & 0--2     & $-2.727\pm 0.182$ & $0.329\pm 0.125$ & 0--2     & $-2.762\pm 0.102$ & $0.275\pm 0.072$\\
 \ion{Ca}{i}  & 6--16    & $-2.602\pm 0.032$ & $0.061\pm 0.017$ & 7--15    & $-2.621\pm 0.020$ & $0.043\pm 0.011$\\
 \ion{Sc}{ii} & 1--4     & $-3.085\pm 0.130$ & $0.105\pm 0.088$ & 1--4     & $-3.091\pm 0.112$ & $0.106\pm 0.083$\\
 \ion{Ti}{i}  & 5--17    & $-2.559\pm 0.037$ & $0.076\pm 0.025$ & 5--16    & $-2.587\pm 0.027$ & $0.048\pm 0.015$\\
 \ion{Ti}{ii} & 10--22   & $-2.564\pm 0.040$ & $0.094\pm 0.019$ & 10--23   & $-2.590\pm 0.028$ & $0.067\pm 0.014$\\
 \ion{V}{i}   & 2--5     & $-3.029\pm 0.062$ & $0.056\pm 0.047$ & 2--6     & $-3.031\pm 0.039$ & $0.029\pm 0.026$\\
 \ion{V}{ii}  & 0--1     & $-2.453\pm 0.093$ &  \dots           & 0--1     & $-2.531\pm 0.067$ &  \dots          \\
 \ion{Cr}{i}  & 2--4     & $-3.011\pm 0.065$ & $0.046\pm 0.033$ & 2--4     & $-3.015\pm 0.047$ & $0.034\pm 0.026$\\
 \ion{Cr}{ii} & 0--1     & $-2.591\pm 0.198$ & \dots            & 0--1     & $-2.719\pm 0.127$ & \dots           \\
 \ion{Mn}{i}  & 2--10    & $-3.025\pm 0.047$ & $0.068\pm 0.029$ & 2--9     & $-3.036\pm 0.029$ & $0.044\pm 0.020$\\
 \ion{Mn}{ii} & 0--1     & $-3.067\pm 0.118$ & \dots            & 0--1     & $-3.059\pm 0.081$ & \dots           \\
 \ion{Fe}{i}  & 38--64   & $-2.996\pm 0.018$ & $0.078\pm 0.011$ & 38--63   & $-3.019\pm 0.012$ & $0.053\pm 0.008$\\
 \ion{Fe}{ii} & 2--6     & $-2.928\pm 0.076$ & $0.101\pm 0.045$ & 2--6     & $-2.955\pm 0.057$ & $0.071\pm 0.032$\\
 \ion{Co}{i}  & 1--8     & $-3.044\pm 0.058$ & $0.071\pm 0.045$ & 1--7     & $-3.047\pm 0.037$ & $0.047\pm 0.275$\\
 \ion{Ni}{i}  & 1--5     & $-2.930\pm 0.064$ & $0.048\pm 0.043$ & 1--5     & $-2.972\pm 0.042$ & $0.029\pm 0.024$\\
 \ion{Sr}{ii} & 0--1     & $-3.093\pm 0.104$ & \dots            & 0--1     & $-3.091\pm 0.073$ & \dots           \\
 \ion{Y}{ii}  & 0--1     & $-3.023\pm 0.117$ & \dots            & 0--1     & $-3.019\pm 0.082$ & \dots           \\
 \ion{Zr}{ii} & 0--1     & $-2.891\pm 0.106$ & \dots            & 0--1     & $-2.939\pm 0.071$ & \dots           \\
 \ion{Ba}{ii} & 1--2     & $-3.099\pm 0.107$ & $0.068\pm 0.059$ & 2         & $-3.106\pm 0.078$ & $0.051\pm 0.039$\\
 \ion{La}{ii} & 0--4     & $-2.792\pm 0.170$ & $0.127\pm 0.154$ & 0--4     & $-2.874\pm 0.122$ & $0.090\pm 0.114$\\
 \ion{Ce}{ii} & 0--1     & $-2.608\pm 0.122$ & \dots            & 0--1     & $-2.651\pm 0.091$ & \dots           \\
 \ion{Nd}{ii} & 0--1     & $-2.903\pm 0.107$ & \dots            & 0--1     & $-2.964\pm 0.070$ & \dots           \\
 \ion{Eu}{ii} & 0--1     & $-2.995\pm 0.084$ & \dots            & 0--1     & $-3.024\pm 0.062$ & \dots           \\                  
\hline                                                
\end{tabular} 
\label{tab:mc_gmp}
\end{center}
\end{table*}

\begin{table}     
\begin{center}       
\caption{Abundances results and uncertainties derived from 1\,000 Monte-Carlo simulations for the (4500/2.0/$0.0$) model.}   
\begin{tabular}{l c r c}                                
\hline
 Element & N lines & $\langle\left[{\rm X/H}\right]\rangle$ & $\sigma$\\
\hline                                                
\hline                                                
 \ion{O}{i}   & 1     & $-0.085\pm 0.083$ & \dots  \\                 
 \ion{Na}{i}  & 0--2     & $-0.101\pm 0.043$ & $0.050\pm 0.036$\\ 
 \ion{Mg}{i}  & 1--2     & $ 0.032\pm 0.028$ & $0.017\pm 0.014$\\ 
 \ion{Al}{i}  & 2--2     & $-0.028\pm 0.055$ & $0.036\pm 0.027$\\ 
 \ion{Si}{i}  & 4--11    & $-0.071\pm 0.038$ & $0.061\pm 0.019$\\ 
 \ion{Ca}{i}  & 3--9     & $-0.018\pm 0.027$ & $0.039\pm 0.014$\\ 
 \ion{Sc}{i}  & 0--1     & $-0.116\pm 0.054$ & \dots  \\                 
 \ion{Sc}{ii} & 2--5     & $-0.017\pm 0.046$ & $0.048\pm 0.024$\\ 
 \ion{Ti}{i}  & 8--19    & $-0.089\pm 0.025$ & $0.056\pm 0.012$\\ 
 \ion{Ti}{ii} & 1     & $-0.088\pm 0.112$ & \dots  \\                 
 \ion{V}{i}   & 5--15    & $-0.081\pm 0.018$ & $0.033\pm 0.009$\\ 
 \ion{Cr}{i}  & 1--2     & $-0.072\pm 0.043$ & $0.053\pm 0.039$\\ 
 \ion{Mn}{i}  & 0--2     & $-0.330\pm 0.084$ & $0.063\pm 0.043$\\ 
 \ion{Fe}{i}  & 41--64   & $-0.065\pm 0.015$ & $0.064\pm 0.007$\\ 
 \ion{Fe}{ii} & 2--4     & $-0.042\pm 0.062$ & $0.070\pm 0.037$\\ 
 \ion{Co}{i}  & 4--10    & $-0.042\pm 0.022$ & $0.040\pm 0.013$\\ 
 \ion{Ni}{i}  & 7--15    & $-0.069\pm 0.028$ & $0.056\pm 0.014$\\ 
 \ion{Y}{ii}  & 0--1     & $-0.021\pm 0.045$ &  \dots \\
 \ion{Zr}{i}  & 2--4     & $-0.089\pm 0.047$ & $0.027\pm 0.020$\\
 \ion{Ba}{ii} & 0--1     & $-0.051\pm 0.083$ & \dots  \\
 \ion{La}{ii} & 2--3     & $-0.026\pm 0.066$ & $0.028\pm 0.023$\\
 \ion{Pr}{ii} & 1      & $-0.006\pm 0.107$ & \dots  \\
\hline                                                
\end{tabular} 
\label{tab:mc_gmr}
\end{center}
\end{table}

\subsubsection{Precision on abundances as a function of signal-to-noise ratio}

\begin{figure}[htb]
   \includegraphics[width=8cm,clip=true]{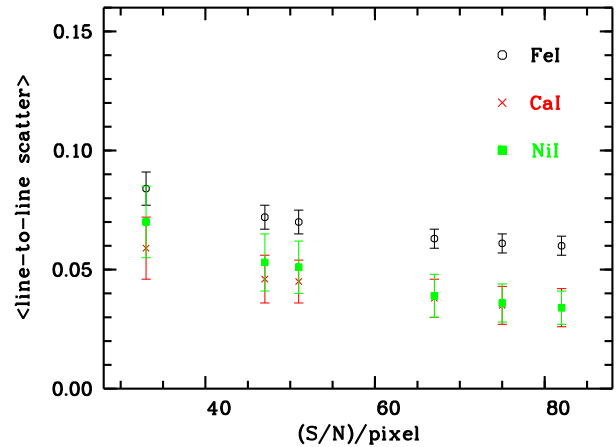} 
 \caption{The average line-to-line scatter for three exemplary elements with a large number of measurable lines, as a function
of the $S/N$ ratio per pixel at the centre of the blue range.}
    \label{snr_sig}
\end{figure}

\begin{figure}[htb] 
   \includegraphics[width=8cm,clip=true]{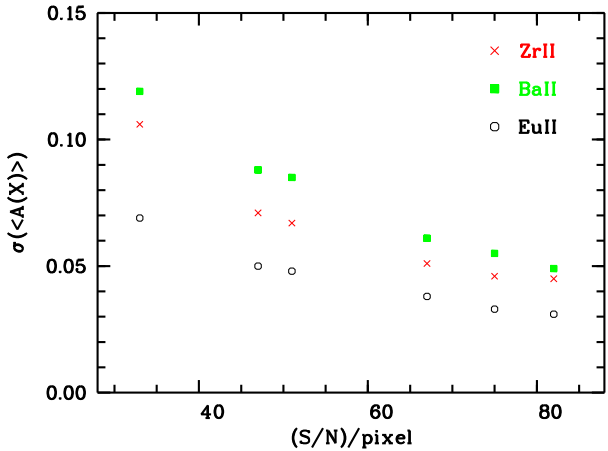} 
 \caption{The scatter on the 1\,000 Monte Carlo events on the 
average abundance for three representative elements with only few available lines, as a function
of the $S/N$ ratio per pixel at the centre of the blue range.}
    \label{snr_save}
\end{figure}

In the next step, we tested the precision we can achieve in the chemical abundance determination as a function of the $S/N$ ratio.
We concentrated on the solar dwarf case (5500/4.0/0.0, $r=14$\,mag), and run 1\,000 Monte Carlo simulations with
six values of exposure time (1\,200, 2\,400, 3\,200, 4\,800, 6\,000, and 7\,200\,s). 
The $S/N$ ratios per pixel, for three different wavelengths on either CCD, in the case
of exposure time of 3\,200\,s for $r=14$, are given in Table\,\ref{tab:snr}.
The results are summarised in Fig.\,\ref{snr_sig} and \ref{snr_save}
for the cases of Fe, Ca, Ni (Fig.~2), and Zr, Ba, Eu (Fig.~3), respectively.

For the elements in Fig.\,\ref{snr_sig}, the abundance is based on several lines, so that the line-to-line
scatter has a meaning. 
For the elements in Fig.\,\ref{snr_save}  
only very few lines are used for the abundance
determination, so that the scatter on the average value over the 1\,000 Monte Carlo events
is  a more robust indicator of the error than the line-to-line scatter. 
The $S/N$ ratio in the figures  is derived in the centre of the blue CCD. 
The relations between the $S/N$ ratios in the various positions on the CCDs  
depends on the exact instrument design, and in this specific case we refer to the phase A 4MOST design (Mignot et al. 2012).
As expected, the line-to-line scatter decreases as the $S/N$ ratio increases;  
the scatter on the average abundance and on the line-to-line scatter
decreases also as the $S/N$ increases.

\begin{table}[htb]                                                                                  
\begin{center}                                                                                     
\caption{Signal-to-noise ratio of the simulated spectra of $r=14$ and exposure time
of 3\,600\,s, at three wavelengths for
both CCDs: 400, 420, and 450\,nm for the blue and 590, 620, and 670\,nm for the red CCD. 
}   
\begin{tabular}{l l l l l l l }                                
\hline                                                       
 Spectrum & \multicolumn{6}{c}{$S/N$ per pixel}\\
 & 400  & 420 & 450 & 590 & 620 & 670 \\
\hline                                                
4500/2.0/0.0   & 33 & 37 & 30 & 70 & 77 & 50\\
4500/2.0/--3.0 & 29 & 37 & 28 & 71 & 79 & 52\\
5500/4.0/0.0   & 44 & 51 & 36 & 72 & 78 & 49\\
5500/4.0/--2.0 & 42 & 50 & 35 & 72 & 79 & 49\\
\hline                                                
\end{tabular} 
\label{tab:snr} 
\end{center}
\end{table}

We then fit a second order polynomial on the line-to-line scatter as a function
of  $S/N$ for those elements, for which the abundance ratios 
were  derived from at least six  lines (\ion{Si}{i}, \ion{Ca}{i}, \ion{Ti}{i}, \ion{V}{i}, \ion{Fe}{i}, \ion{Fe}{ii}, \ion{Ni}{i}). 
The results are shown in Table\,\ref{tab:poly_fit}. 
These values can be interpolated to reach 
the practical  request that, {\em in order for  the line-to-line scatter to be smaller than 0.1\,dex for all elements, 
a $S/N\ge 50$ per pixel at the centre of the blue CCD needs to be reached.}

For the cases where only  a single line, or few lines, are detectable in the spectral range, the r.m.s. scatter on the average
value is a better measure for the abundance precision, as argued above. 
In analogy to above, we  fitted a second order polynomial to the scatter on the average A(X), as derived from the Monte Carlo events, vs. $S/N$ ratio.
These results are shown in Table\,\ref{tab:poly_fit} for all species.
Hence, the constraint $S/N=50$ per pixel  (again, in the centre of the blue CCD)  
for all species will yield $\sigma _{\langle \rm A(X)\rangle} < 0.1$\,dex, except for \ion{Cr}{ii} and \ion{La}{ii} for which
we find a limiting precision of 0.11 and 0.12\,dex, respectively.

\begin{table*}[htb]                                                                                  
\begin{center}                                                                                     
\caption{Coefficients of a 2nd order polynomial fit to the line-to-line scatter and
$\sigma _{\rm average}$ as a function of S/N ratio, for the case 5500/4.0/0.0.}  
\begin{tabular}{l | l l l | l l l }                                
\hline                                                       
Element & \multicolumn{3}{c}{line-to-line scatter} & \multicolumn{3}{c}{$\sigma _{\rm average}$}\\
        & $a_0$ & $a_1$ & $a_2$ & $a_0$ & $a_1$ & $a_2$ \\
\hline                                                
\ion{Na}{i}  &                        &                        &                       & $0.137$               & $-0.224\times 10^{-2}$ & $0.135\times 10^{-4}$\\
\ion{Al}{i}  &                        &                        &                       & $0.144$               & $-0.253\times 10^{-2}$ & $0.146\times 10^{-4}$\\
\ion{Mg}{i}  &                        &                        &                       & $0.908\times 10^{-1}$ & $-0.153\times 10^{-2}$ & $0.864\times 10^{-5}$\\
\ion{Si}{i}  & $ 0.121$               & $-0.188\times 10^{-2}$ & $0.109\times 10^{-4}$ & $0.592\times 10^{-1}$ & $-0.850\times 10^{-3}$ & $0.424\times 10^{-5}$\\
\ion{Si}{ii} &                        &                        &                       & $0.190$               & $-0.341\times 10^{-2}$ & $0.207\times 10^{-4}$\\
\ion{Ca}{i}  & $ 0.993\times 10^{-1}$ & $-0.153\times 10^{-2}$ & $0.899\times 10^{-5}$ & $0.533\times 10^{-1}$ & $-0.862\times 10^{-3}$ & $0.471\times 10^{-5}$\\
\ion{Sc}{ii} &                        &                        &                       & $0.133$               & $-0.212\times 10^{-2}$ & $0.111\times 10^{-4}$\\
\ion{Ti}{i}  & $ 0.144$               & $-0.243\times 10^{-2}$ & $0.151\times 10^{-4}$ & $0.619\times 10^{-1}$ & $-0.103\times 10^{-2}$ & $0.592\times 10^{-5}$\\
\ion{Ti}{ii} &                        &                        &                       & $0.131$               & $-0.257\times 10^{-2}$ & $0.163\times 10^{-4}$\\
\ion{V}{i}   & $ 0.902\times 10^{-1}$ & $-0.456\times 10^{-3}$ & $0.132\times 10^{-5}$ & $0.576\times 10^{-1}$ & $-0.106\times 10^{-2}$ & $0.650\times 10^{-5}$\\
\ion{Cr}{i}  &                        &                        &                       & $0.142$               & $-0.264\times 10^{-2}$ & $0.159\times 10^{-4}$\\
\ion{Cr}{ii} &                        &                        &                       & $0.318$               & $-0.583\times 10^{-2}$ & $0.338\times 10^{-4}$\\
\ion{Mn}{i}  &                        &                        &                       & $0.734\times 10^{-1}$ & $-0.906\times 10^{-3}$ & $0.410\times 10^{-5}$\\
\ion{Fe}{i}  & $ 0.125$               & $-0.156\times 10^{-2}$ & $0.934\times 10^{-5}$ & $0.195\times 10^{-1}$ & $-0.276\times 10^{-3}$ & $0.151\times 10^{-5}$\\
\ion{Fe}{ii} & $ 0.187$               & $-0.248\times 10^{-2}$ & $0.155\times 10^{-4}$ & $0.911\times 10^{-1}$ & $-0.153\times 10^{-2}$ & $0.855\times 10^{-5}$\\
\ion{Co}{i}  &                        &                        &                       & $0.118$               & $-0.225\times 10^{-2}$ & $0.138\times 10^{-4}$\\
\ion{Ni}{ii} & $ 0.126$               & $-0.207\times 10^{-2}$ & $0.117\times 10^{-4}$ & $0.577\times 10^{-1}$ & $-0.886\times 10^{-3}$ & $0.438\times 10^{-5}$\\
\ion{Y}{ii}  &                        &                        &                       & $0.182$               & $-0.320\times 10^{-2}$ & $0.191\times 10^{-4}$\\
\ion{Zr}{ii} &                        &                        &                       & $0.225$               & $-0.462\times 10^{-2}$ & $0.297\times 10^{-4}$\\
\ion{Ba}{ii} &                        &                        &                       & $0.214$               & $-0.348\times 10^{-2}$ & $0.180\times 10^{-4}$\\
\ion{La}{ii} &                        &                        &                       & $0.300$               & $-0.513\times 10^{-2}$ & $0.288\times 10^{-4}$\\
\ion{Nd}{ii} &                        &                        &                       & $0.221$               & $-0.365\times 10^{-2}$ & $0.210\times 10^{-4}$\\
\ion{Eu}{ii} &                        &                        &                       & $0.127$               & $-0.218\times 10^{-2}$ & $0.125\times 10^{-4}$\\
\hline                                                
\end{tabular} 
\label{tab:poly_fit} 
\end{center}
\end{table*}

%
%
\subsection{Radial velocity precision}

We determined radial velocities by a cross-correlation of our noise-injected 
synthetic spectra, for different values of resolution and $S/N$ ratio, 
against the synthetic spectra as templates. This study does not take into account
any poor knowledge on atomic data (laboratory wavelengths and oscillator strength), 
possible mismatches between the stellar
spectrum and the cross-correlation mask, nor zero-point considerations. 
The mismatch between the observed spectrum and the template will mostly depend  on the type of star. Figure\,1 of Nidever et al. (2002) 
illustrates this effect of the   precision on radial velocity. 
For instance, measurements made with a Solar template 
lead to an mismatch, thus error,  that is increasing much faster towards F-type stars than towards K-type stars. 
The mismatch has also a greater effect for hot stars (Verschueren, David, \& Vrancken 1999).
The zero point is directly dependent on the accuracy of the instrumental wavelength calibration and/or 
the manner in which it is corrected for using standard sources (stars, asteroids, telluric rays, ...).

In practice,  we pursued three different approaches, the results of which are listed in Table\,\ref{tlab} and which are labelled as:
(i) ``Gauss'':  strong lines ({\ion{Ca}{ii}-K and H;  H$\gamma$ and H$\delta$}) were masked out and the cross correlation function (CCF) peak was 
 fitted with a Gaussian;  
(ii) ``PolyM'': strong lines are masked out as above, but the CCF
peak was fitted with a third-order polynomial; 
(iii) ``Poly'': no line-masking has been applied and the CCF
peak was determined by third-order polynomial fit as in (ii). 
 
\begin{table*}[htb]
\caption{Radial velocity errors. The two ranges are similar to our selected blue range but in one case \ion{Ca}{ii}-K
is in, in the other not, to check the impact of this strong line in the radial velocity determination.} 
\label{tlab}
\begin{tabular}{cccccccccc}
\hline
& & & \multicolumn{7}{c}{$\sigma$v [km\,s$^{-1}$]} \\ 
\raisebox{1.5ex}[-1.5ex]{T$_{\rm eff}$  / log\,$g$ / / [$\alpha$/Fe] / [M/H] }  & $R$ & \raisebox{1.5ex}[-1.5ex]{$S/N$} & \multicolumn{3}{c}{(390--450\,nm)} && \multicolumn{3}{c}{(400--460\,nm)}\\
\cline{4-6} \cline{8-10}
\raisebox{1.5ex}[-1.5ex]{$[$K$]$ / [cgs] / [km\,s$^{-1}$] / [dex] / [dex]} & & \raisebox{1.5ex}[-1.5ex]{[pixel$^{-1}$]} & Gauss & PolyM & Poly && Gauss & PolyM & Poly \\
\hline
4700 / 2.0 / 0.4 / $-$2.0 & 15\,000 & 100 & 0.022 & 0.015 & 0.014 && 0.022 & 0.015 & 0.014 \\  
4700 / 2.0 / 0.4 / $-$2.0 & 15\,000 &  50 & 0.044 & 0.030 & 0.027 && 0.041 & 0.029 & 0.027 \\  
4700 / 2.0 / 0.4 / $-$2.0 & 15\,000 &  25 & 0.084 & 0.057 & 0.052 && 0.089 & 0.062 & 0.056 \\  
4700 / 2.0 / 0.4 / $-$2.0 & 20\,000 & 100 & 0.016 & 0.011 & 0.010 && 0.016 & 0.011 & 0.011 \\  
4700 / 2.0 / 0.4 / $-$2.0 & 20\,000 &  50 & 0.031 & 0.022 & 0.021 && 0.031 & 0.022 & 0.022 \\  
4700 / 2.0 / 0.4 / $-$2.0 & 20\,000 &  25 & 0.061 & 0.044 & 0.041 && 0.062 & 0.044 & 0.041 \\  
4700 / 2.0 / 0.4 / $-$2.0 & 25\,000 & 100 & 0.013 & 0.010 & 0.009 && 0.013 & 0.009 & 0.009 \\  
4700 / 2.0 / 0.4 / $-$2.0 & 25\,000 &  50 & 0.025 & 0.019 & 0.018 && 0.025 & 0.019 & 0.018 \\  
4700 / 2.0 / 0.4 / $-$2.0 & 25\,000 &  25 & 0.050 & 0.037 & 0.034 && 0.048 & 0.036 & 0.035 \\ 
\hline	      						    
4750 / 2.0 / 0.0 / \phantom{$-$}0.0  & 15\,000 & 100 & 0.013 & 0.009 & 0.009 && 0.013 & 0.009 & 0.009 \\  
4750 / 2.0 / 0.0 / \phantom{$-$}0.0  & 15\,000 &  50 & 0.026 & 0.019 & 0.018 && 0.026 & 0.019 & 0.017 \\  
4750 / 2.0 / 0.0 / \phantom{$-$}0.0  & 15\,000 &  25 & 0.052 & 0.037 & 0.034 && 0.051 & 0.036 & 0.035 \\  
4750 / 2.0 / 0.0 / \phantom{$-$}0.0  & 20\,000 & 100 & 0.009 & 0.007 & 0.007& & 0.009 & 0.007 & 0.006 \\  
4750 / 2.0 / 0.0 / \phantom{$-$}0.0  & 20\,000 &  50 & 0.019 & 0.014 & 0.013 && 0.019 & 0.014 & 0.013 \\  
4750 / 2.0 / 0.0 / \phantom{$-$}0.0  & 20\,000 &  25 & 0.037 & 0.028 & 0.025 && 0.038 & 0.027 & 0.025 \\  
4750 / 2.0 / 0.0 / \phantom{$-$}0.0  & 25\,000 & 100 & 0.008 & 0.006 & 0.005 && 0.008 & 0.006 & 0.005 \\  
4750 / 2.0 / 0.0 / \phantom{$-$}0.0  & 25\,000 &  50 & 0.015 & 0.011 & 0.011 && 0.015 & 0.011 & 0.011 \\  
4750 / 2.0 / 0.0 / \phantom{$-$}0.0  & 25\,000 &  25 & 0.030 & 0.023 & 0.021 && 0.031 & 0.023 & 0.021 \\
\hline		       
\end{tabular}	 
\end{table*}

We cannot see any tangible difference between the two spectral ranges we probed; thus we conclude
that the \ion{Ca}{ii}-K and -H lines do not affect the radial velocity
determination  in any way.
The precision in radial velocity determined here is much higher than the one expected from Gaia
(see Katz at al. 2004) 
for the entire sample of 4MOST stars. 
The reason is that the high resolution of 4MOST is larger than the one of Gaia. 
Furthermore, the blue range of the 4MOST spectrograph contains a wealth of spectral lines, even in  low
metallicity stars, which facilitates the derivation of radial velocity at high precision.

Bearing in mind that we are aiming at 
a precision of 1\kms, these results suggest
that the limitation to radial velocity precision
does not stem from $S/N$ ratio, resolution
or spectral coverage, as the errors in radial
velocity in our tests were always  better than this specification by over
an order of magnitude.
In practice, we expect constraints to rather come
from other 
technical limitations of the spectrograph
(mechanical and  thermal instabilities, photocentre motion, scrambling achieved by the fibre,
fibre cross-talk).
If such aspects can be kept under control,
our results indicate that a precision in radial velocity of better than 1\kms\
should be achievable. 
In all cases, the unmasked parabolic fits yield the smallest uncertainties. 

A consequence of this analysis is that, as long as we aim
at a precision of the order of 1\kms,  
the wavelength range and resolution of a high-resolution spectrograph should be rather driven 
by the optimisation of  the chemical analysis,
as  elaborated in Section 5.3.

%
\section{Summary}
With accurate information for 1\% of Galactic stars, Gaia
will allow a breakthrough in our understanding of the MW. But this new
knowledge would not be complete without the additional information
that can be provided by high resolution spectroscopy. Anticipating the first data release
of Gaia, and in the perspective of exploiting optimally our new insights on the Milky Way, there is
a primary need for a follow-up ground-based spectroscopic survey. With this in mind, 
we explored the spectral quality such an instrument needs to deliver to achieve meaningful 
and accurate scientific results on stellar abundances.

Based on extensive tests carried out on synthetic spectra over a broad range of 
stellar parameters we argue for the following, optimal and yet technically feasible 
design:
 
\begin{itemize}
\item{ } {\em Wavelength ranges}:\\
blue arm: 
[395;459]\,nm, 
with eventually any gap between detectors avoiding to split the G-band , [429;432]\,nm,
in the two detectors;\\ 
red arm: [587;673]\,nm, with, if necessary,
a gap around 635\,nm.
\item{ }{\em Resolving Power:} 20\,000 on average across the above wavelength range, but never smaller than 18\,000.
\item{ }{\em Pixel-Size} is not a strong limitation, and we suggest to have
 an optimal sampling with two elements. As good compromises  we envisage as values 
 for Resolving Power/Sampling: 25\,000/2.0--2.5, 20\,000/2.5, 15\,000/2.5--3.0. 
\end{itemize}

An instrument with such a design will allow to derive abundances
for the elements listed in Table\,\ref{range}.
It can provide a precision in the abundance
determination of about 0.1\,dex, for the majority of the elements.
Tables \ref{resps}, \ref{oitab}, and \ref{euiitab}
can be used to estimate the typical accuracies from the single lines
as a function of line strength and $S/N$ ratio.

None of the above has a major influence on the reachable precision on {\em radial velocity}. 
Moreover, this is limited by 
systematics in the instrument design such as mechanical stability, which is beyond the 
scope of the present spectral feasibility study.

\acknowledgements
We thank the 4MOST consortium members for making their preliminary performance numbers available to us, 
which enabled a realistic assessment of the spectral quality from a 4\,m MOS instrument.
AK thanks the Deutsche Forschungsgemeinschaft for funding through Emmy-Noether grant  Ko 4161/1. 
EC, LS, CJH, NC, HGL, and EKG acknowledge financial support by the Sonderforschungsbereich SFB\,881
``The Milky Way System'' (subprojects A4 and A5) of the German Research Foundation (DFG).
PB acknowledge support from the Programme National
de Physique Stellaire (PNPS) and the Programme National
de Cosmologie et Galaxies (PNCG) of the Institut National de Sciences
de l'Universe of CNRS.
AH acknowledges funding support from the European 
Research Council under ERC-StG grant GALACTICA-240271.
\appendix
\section{Relation of pixel vs. line strength}
In what follows, we represent an absorption line as a Gaussian profile:
$$F(\lambda) = 1 - A e^{-{\left({\lambda - \lambda _\circ}\right)^2\over {2\sigma ^2}}}.$$ 

We take the relative depression of the line core $A$ as a measure
of the line strength.
We consider a line as {\em well-defined} when it covers  a certain number of pixels from the core to the continuum.
In particular, we parameterise the extent of the line by chosing a suitable (small) number $\delta$ 
such that the feature transits into the continuum if  $F > 1-\delta$. Obviously, 
$\delta$ is dependent on the $S/N$ level, but it is typically of the order of  
$\delta =$0.01 for continuum normalized spectra.

Here we are interested in determining a  relation between the line strength, $A$,  and the number of
pixels it covers. From the above relation
$$ 1 - A e^{-{\left({\lambda - \lambda _\circ}\right)^2\over {2\sigma ^2}}} < 1-\delta$$
it follows that 
$$\left({\lambda - \lambda _\circ}\right)^2 < 2\sigma ^2\ln{A\over\delta}, $$ 
which results in a full width of the line of
$$2|\lambda _1 -\lambda _\circ| = 2\sigma\sqrt{2\ln{A\over\delta}},$$
where $\lambda_1$ is the wavelength where the line profiles transits into the continuum.

The number of pixel defining the line profile is thus:
$$N = 2{\sigma \sqrt{2\ln{A\over\delta}}\over{\rm PixelSize}}.$$

Now we intend to make a connection between  $\sigma$ and spectral resolving
power given by the FWHM of the spectrograph's instrumental profile $\Delta\lambda_\mathrm{FWHM}$.
At the resolution of 4MOST, stellar lines are typically not resolved, and the
observed line profile largely reflects the instrumental profile so that

$$\sigma \approx \frac{\Delta\lambda_\mathrm{FWHM}}{2\sqrt{2\ln 2}}
\approx \frac{\Delta\lambda_\mathrm{FWHM}}{2.355}.$$

If we introduce the Sampling, $S$,  as $S = {\Delta\lambda_\mathrm{FWHM}\over {\rm PixelSize}}$
we obtain
$$N \approx S\sqrt{\ln\frac{A}{\delta}/{\ln 2}}\approx 1.2\times S\sqrt{\ln\frac{A}{\delta}}.$$
The number of pixels that define the line is proportional to the square root of the
logarithm of the strength of the line itself.

For a sampling of 2.0, a strong line with a residual intensity of 0.1 is thus defined by only $\sim$5 pixels,
while a weak line of residual intensity 0.9 covers less than 4 pixels.
With the 2.5 sampling of the centre of the ranges, strong and weak lines are
defined by more than 6 and less than 5 pixels, respectively.
Likewise, at a sampling of 3.0 these numbers become less than 8 and more than 5, respectively. 
This should be considered  a lower limit, however,  since strong lines are better described with a wider Voigt profile.


\begin{thebibliography}{}
%
\bibitem[Arlandini et al.(1999)]{1999ApJ...525..886A} Arlandini, C., K{\"a}ppeler, F., Wisshak, K., et al.:\ 1999, \apj\ 525, 886 
%
\bibitem[Babusiaux et al.(2010)]{2010A&A...519A..77B} Babusiaux, C., G{\'o}mez, A., Hill, V., et al.:\ 2010, A\&A 519, A77 
%
\bibitem[Bailer-Jones(2002)]{2002Ap&SS.280...21B} Bailer-Jones, C.~A.~L.:\ 2002, Ap\&SS, 280, 21 
%
\bibitem[Balcells et al.(2010)]{2010SPIE.7735E.242B} Balcells, M., Benn, C.~R., Carter, D., et al.:\ 2010, SPIE 7735,  
%
\bibitem[Barden et al.(2010)]{2010SPIE.7735E...8B} Barden, S.~C., Jones, D.~J., Barnes, S.~I., et al.:\ 2010, SPIE 7735,  
%
\bibitem[Barbuy(1983)]{1983A&A...123....1B} Barbuy, B.:\ 1983, \aap\ 123, 1
%
\bibitem[Beers et al.(1985)]{1985AJ.....90.2089B} Beers, T.~C., Preston, G.~W., \& Shectman, S.~A.:\ 1985, \aj\ 90, 2089 
%
\bibitem[Beers \& Christlieb(2005)]{2005ARA&A..43..531B} Beers, T.~C., \& Christlieb, N.:\ 2005, \araa\ 43, 531 
%
\bibitem[Bensby et al.(2005)]{2005A&A...433..185B} Bensby, T., Feltzing, S., Lundstr{\"o}m, I., \& Ilyin, I.:\ 2005, A\&A 433, 185 
\bibitem[Bensby et al.(2010)]{2010A&A...512A..41B} Bensby, T., Feltzing, S., Johnson, J.~A., et al.:\ 2010, A\&A 512, A41 
%
\bibitem[Bensby et al.(2011)]{2011A&A...533A.134B} Bensby, T., Ad{\'e}n, D., Mel{\'e}ndez, J., et al.:\ 2011, \aap\ 533, A134
%
\bibitem[Bernstein et al.(2003)]{2003SPIE.4841.1694B} Bernstein, R., 
Shectman, S.~A., Gunnels, S.~M., Mochnacki, S., 
\& Athey, A.~E.\ 2003, \procspie, 4841, 1694 
%
\bibitem[Binney et al.(1997)]{binney} Binney, J., Gerhard, O., \& Spergel, D.:\ 1997, \mnras\ 288, 365 
%
\bibitem[Bland-Hawthorn et al.(2010)]{2010ApJ...713..166B} Bland-Hawthorn, J., Krumholz, M.~R., \& Freeman, K.:\ 2010, \apj\ 713, 166
%
\bibitem[Bonifacio et al.(2009)]{bonifacio09} Bonifacio, P., Spite, M., Cayrel, R., et al.:\ 2009, A\&A 501, 519 
%
\bibitem[Bonifacio et al.(2010)]{Bonifacio2010SPIE} Bonifacio, P., Arenou, F., Babusiaux, C., et al.:\ 2010, \procspie\ 7735,  
%
\bibitem[Bonifacio et al.(2011)]{2011EAS....45..219B} Bonifacio, P., Mignot, S., Dournaux, J.-L., et al.:\ 2011, EAS Publications Series 45, 219 
%
\bibitem[Bovy et al.(2012)]{2012ApJ...751..131B} Bovy, J., Rix, H.-W., \& Hogg, D.~W.:\ 2012, \apj\ 751, 131 
%
\bibitem[Caffau et al.(2008)]{caffau08} Caffau, E., Ludwig, H.-G., Steffen, M., et al.:\ 2008, A\&A 488, 1031
%
\bibitem[Caffau et al.(2011)]{2011Natur.477...67C} Caffau, E., Bonifacio, P., Fran{\c c}ois, P., et al.:\ 2011a, Nature 477, 67 
%
\bibitem[Caffau et al.(2011)]{2011SoPh..268..255C} Caffau, E., Ludwig, H.-G., Steffen, M., Freytag, B., \& Bonifacio, P.:\ 2011b, Sol. Phys. 268, 255 
%
\bibitem[Carollo et al.(2007)]{2007Natur.450.1020C} Carollo, D., Beers, T.~C., Lee, Y.~S., et al.:\ 2007, \nat\ 450, 1020
%
\bibitem[Castelli \& Kurucz (2003)]{castelli03} Castelli, F., \& Kurucz, R. L.:\ 2003, Proceed. of IAU Symp. 210, Modelling of Stellar Atmospheres, eds. N. Piskunov et al., poster A20 on the enclosed CD-ROM ({\tt astro-ph/0405087})
%
\bibitem[Cayrel(1988)]{1988IAUS..132..345C} Cayrel, R.:\ 1988, The Impact of Very High S/N Spectroscopy on Stellar Physics,
Proceedings of the 132nd Symposium of IAU, Edited by G. Cayrel de Strobel and Monique Spite, 132, 345 
%
\bibitem[Cayrel et al.(2004)]{cayrel04} Cayrel, R., Depagne, E., Spite, M., et al.:\ 2004, A\&A 416, 1117 
%
\bibitem[Chiappini et al.(2001)]{2001ApJ...554.1044C} Chiappini, C., Matteucci, F., \& Romano, D.\ 2001, \apj, 554, 1044 
%
\bibitem[Chiappini(2011)]{2011EAS....45..293C} Chiappini, C.:\ 2011, EAS Publications Series 45, 293 
%
\bibitem[Christlieb et al.(2004)]{2004ApJ...603..708C} Christlieb, N.,  Gustafsson, B., Korn, A.~J., et al.:\ 2004, \apj\ 603, 708 
%
\bibitem[Cirasuolo et al.(2011)]{2011Msngr.145...11C} Cirasuolo, M., Afonso, J., Bender, R., et al.:\ 2011, The Messenger 145, 11 
%
\bibitem[Combes et al.(1990)]{1990A&A...233...82C} Combes, F., Debbasch, F., Friedli, D., \& Pfenniger, D.:\ 1990, A\&A 233, 82 
%
\bibitem[D{\'{\i}}az et al.(2011)]{2011A&A...531A.143D} D{\'{\i}}az, C.~G., Gonz{\'a}lez, J.~F., Levato, H., \& Grosso, M.:\ 2011, \aap\ 531, A143 
%
\bibitem[Dierickx et al.(2010)]{2010ApJ...725L.186D} Dierickx, M., Klement, R., Rix, H.-W., \& Liu, C.:\ 2010, \apjl\ 725, L186 
%
\bibitem[Dwek et al.(1995)]{dwek} Dwek, E., Arendt, R.~G., Hauser, M.~G., et al.:\ 1995, \apj\ 445, 716 
%
\bibitem[Edvardsson et al.(1993)]{1993A&A...275..101E} Edvardsson, B., Andersen, J., Gustafsson, B., et al.:\ 1993, \aap\ 275, 101 
%
\bibitem[Eisenstein et al.(2011)]{2011AJ....142...72E} Eisenstein, D.~J., Weinberg, D.~H., Agol, E., et al.:\ 2011, \aj\ 142, 72 
%
\bibitem[Fabricant et al.(2005)]{2005PASP..117.1411F} Fabricant, D., Fata, 
R., Roll, J., et al.\ 2005, \pasp, 117, 1411 
%
\bibitem[Fran{\c c}ois et al.(2007)]{2007A&A...476..935F} Fran{\c c}ois, P., Depagne, E., Hill, V., et al.:\ 2007, \aap\ 476, 93
%
\bibitem[Frebel et al.(2005)]{2005Natur.434..871F} Frebel, A., Aoki, W., Christlieb, N., et al.:\ 2005, Nature 434, 871 
%
\bibitem[Freeman \& Bland-Hawthorn(2002)]{2002ARA&A..40..487F} Freeman, K., \& Bland-Hawthorn, J.:\ 2002, \araa\ 40, 487 
%
\bibitem[Fulbright et al.(2006)]{2006ApJ...636..821F} Fulbright, J.~P., McWilliam, A., \& Rich, R.~M.\ 2006, \apj, 636, 821 
%
\bibitem[Fulbright et al.(2007)]{2007ApJ...661.1152F} Fulbright, J.~P., McWilliam, A., \& Rich, R.~M.:\ 2007, \apj\ 661, 1152 
%
\bibitem[Fuhrmann(2008)]{2008MNRAS.384..173F} Fuhrmann, K.:\ 2008, \mnras\ 384, 173 
%
\bibitem[Gilmore et al.(2012)]{2012Msngr.147...25G} Gilmore, G., Randich, S., Asplund, M., et al.:\ 2012, The Messenger 147, 25 
%
\bibitem[Gonzalez et al.(2011)]{gonzalez} Gonzalez, O.~A., Rejkuba, M., Zoccali, M., et al.:\ 2011, \aap\ 530, A54 
%
\bibitem[Gray(2005)]{2005oasp.book.....G} Gray, D.~F.:\ 2005, ``The 
Observation and Analysis of Stellar Photospheres'', 3rd Edition, by D.F.~Gray.~ISBN 
0521851866.
UK: Cambridge University Press, 2005
%
\bibitem[Grenon(1999)]{1999Ap&SS.265..331G} Grenon, M.:\ 1999, \apss\ 265, 331 
%
\bibitem[Haywood(2008)]{2008MNRAS.388.1175H} Haywood, M.:\ 2008, \mnras\ 388, 1175 
%
\bibitem[Heil et al.(2009)]{2009PASA...26..243H} Heil, M., Juseviciute, A., K{\"a}ppeler, F., et al.:\ 2009, \pasa\ 26, 243 
%
\bibitem[Hill et al. (2011)]{hill11} Hill, V., Lecureur, A., G{\'o}mez, A., et al.:\ 2011, A\&A 534, A80
%
\bibitem[Howard et al.(2008)]{2008ApJ...688.1060H} Howard, C.~D., Rich, R.~M., Reitzel, D.~B., et al.:\ 2008, \apj\ 688, 1060 
%
\bibitem[Howard et al.(2009)]{2009ApJ...702L.153H} Howard, C.~D., Rich, R.~M., Clarkson, W., et al.:\ 2009, \apjl\ 702, L153 
%
\bibitem[de Jong (2011)]{dejong11} de Jong, R.:\ 2011, The Messenger 145, 14
%
\bibitem[de Jong et al.(2012)]{2012EPJWC..1909004D} de Jong, R.~S., Chiappini, C., \& Schnurr, O.:\ 2012a, Assembling the Puzzle of the Milky Way, Le Grand-Bornand, France, Edited by C.~Reyl{\'e}; A.~Robin; M.~Schultheis; EPJ Web of Conferences, Volume 19, id.09004, 19, 9004 
%
\bibitem[de Jong et al.(2012)]{2012arXiv1206.6885D} de Jong, R.~S., 
Bellido-Tirado, O., Chiappini, C., et al.:\ 2012b, arXiv:1206.6885, to appear on
Proceedings of the SPIE Astronomical Instrumentation + Telescopes conference, Amsterdam, 2012
%
\bibitem[Johnson et al.(2011)]{2011ApJ...732..108J} Johnson, C.~I., Rich, R.~M., Fulbright, J.~P., Valenti, E., \& McWilliam, A.:\ 2011, \apj\ 732, 108 
%
\bibitem[Karlsson et al.(2011)]{2011arXiv1101.4024K} Karlsson, T., Bromm, 
V., \& Bland-Hawthorn, J.\ 2011, arXiv:1101.4024 
%
\bibitem[Katz et al.(2004)]{2004MNRAS.354.1223K} Katz, D., Munari, U., Cropper, M., et al.:\ 2004, \mnras\ 354, 1223 
%
\bibitem[Klessen et al.(2012)]{2012MNRAS.421.3217K} Klessen, R.~S., Glover, S.~C.~O., \& Clark, P.~C.:\ 2012, \mnras\ 421, 3217 
%
\bibitem[Koch \& McWilliam(2008)]{2008AJ....135.1551K} Koch, A., \& McWilliam, A.:\ 2008, \aj\ 135, 1551 
%
\bibitem[Koch(2009)]{2009AN....330..675K} Koch, A.:\ 2009, Astronomische Nachrichten 330, 675 
%
\bibitem[Kormendy \& Kennicutt(2004)]{2004ARA&A..42..603K} Kormendy, J., \& Kennicutt, R.~C., Jr.:\ 2004, \araa\ 42, 603 
%
\bibitem[Kunder et al.(2012)]{2012AJ....143...57K} Kunder, A., Koch, A., Rich, R.~M., et al.:\ 2012, \aj\ 143, 57 
%
\bibitem[Kurucz (2005)]{kurucz05} Kurucz, R.~L.: 2005, Memorie della Societ\`a Astronomica  Italiana Supplementi 8, 14
%
\bibitem[Lee et al.(2008)]{2008AJ....136.2022L} Lee, Y.~S., Beers, T.~C., Sivarani, T., et al.:\ 2008, \aj\ 136, 2022 
%
\bibitem[Lindegren(2010)]{2010IAUS..261..296L} Lindegren, L.:\ 2010, ``Relativity in Fundamental Astronomy'',
Proceedings of the IAU Symposium, Volume 261, p. 296-305
%
\bibitem[Matteucci(2003)]{2003Ap&SS.284..539M} Matteucci, F.:\ 2003, \apss\ 284, 539 
%
\bibitem[McWilliam \& Zoccali(2010)]{2010ApJ...724.1491M} McWilliam, A., \& Zoccali, M.:\ 2010, \apj\ 724, 1491 
%
\bibitem[Melo et al.(2001)]{2001A&A...375..851M} Melo, C.~H.~F., Pasquini, L., \& De Medeiros, J.~R.:\ 2001, \aap\ 375, 851 
%
\bibitem[Minchev \& Famaey(2010)]{2010ApJ...722..112M} Minchev, I., \& Famaey, B.:\ 2010, \apj\ 722, 112 
%
\bibitem[Mignot et al.(2010)]{Mignot} Mignot, S., Cohen, M., Dalton, G., et al.:\ 2010, \procspie\ 7735,  
%
\bibitem[Minchev et al.(2012)]{2012arXiv1205.6475M} Minchev, I., Famaey, 
B., Quillen, A.~C., et al.:\ 2012, arXiv:1205.6475 
%
\bibitem[Monaco et al.(2010)]{M10} Monaco, L., Bonifacio, P., Sbordone, L., Villanova, S., \& Pancino, E.:\ 2010, \aap\ 519, L3 
%
\bibitem[Monaco et al.(2012)]{M12} Monaco, L., Villanova, S., Bonifacio, P., et al.:\ 2012, \aap\ 539, A157 
%
\bibitem[Mucciarelli et al.(2012a)]{Mucciarelli1} Mucciarelli, A., Salaris, M., \& Bonifacio, P.:\ 2012a, \mnras\ 419, 2195 
%
\bibitem[Mucciarelli et al.(2012b)]{Mucciarelli2} Mucciarelli, A., Salaris, M., \& Bonifacio, P.:\ 2012b, MSAIS 22, 86  
%
\bibitem[Munari et al.(2001)]{2001BaltA..10..613M} Munari, U., Agnolin, P., \& Tomasella, L.:\ 2001, Baltic Astronomy v.10, p.613-627., 10, 613 
%
\bibitem[Munari \& Castelli(2000)]{2000A&AS..141..141M} Munari, U., \& Castelli, F.:\ 2000, A\&AS 141, 141 
%
\bibitem[Nidever et al.(2002)]{2002ApJS..141..503N} Nidever, D.~L., Marcy, G.~W., Butler, R.~P., Fischer, D.~A., \& Vogt, S.~S.:\ 2002, \apjs\ 141, 503 
%
\bibitem[Nissen \& Schuster(2010)]{NissenSchuster} Nissen, P.~E., \& Schuster, W.~J.:\ 2010, A\&A 511, L10 
%
\bibitem[Pasquini et al.(2002)]{2002Msngr.110....1P} Pasquini, L., Avila, G., Blecha, A., et al.:\ 2002, The Messenger 110, 1 
%
\bibitem[Perryman \& ESA(1997)]{hipparcos} Perryman, M.~A.~C., \& ESA: 1997, ESA Special Publication, 1200 
%
\bibitem[Perryman et al.(2001)]{2001A&A...369..339P} Perryman, M.~A.~C., de Boer, K.~S., Gilmore, G., et al.:\ 2001, A\&A 369, 339 
%
\bibitem[Piontek \& Steinmetz(2011)]{2011MNRAS.410.2625P} Piontek, F., \& Steinmetz, M.:\ 2011, \mnras\ 410, 2625 
%
\bibitem[Pritzl et al.(2005)]{2005AJ....130.2140P} Pritzl, B.~J., Venn, K.~A., \& Irwin, M.:\ 2005, \aj\ 130, 2140 
%
\bibitem[Reddy et al.(2006)]{2006MNRAS.367.1329R} Reddy, B.~E., Lambert, D.~L., \& Allende Prieto, C.:\ 2006, \mnras\ 367, 1329 
%
\bibitem[Roederer \& Lawler(2012)]{2012ApJ...750...76R} Roederer, I.~U., \& Lawler, J.~E.:\ 2012, \apj\ 750, 76 
%
\bibitem[Ryde et al.(2010)]{2010A&A...509A..20R} Ryde, N., Gustafsson, B., Edvardsson, B., et al.:\ 2010, A\&A 509, A20 
%
\bibitem[Sales et al.(2009)]{2009MNRAS.400L..61S} Sales, L.~V., Helmi, A., Abadi, M.~G., et al.:\ 2009, \mnras\ 400, L61 
%
\bibitem[Sartoretti et al.(2012)]{sartoretti12} Sartoretti, P., Leclerc, N., Walcher, C.J., Caffau, E., \&  Sbordone, L.:\  2012, SPIE in press
%
\bibitem[Sbordone (2005)]{sbordone05} Sbordone, L.: 2005, Memorie della Societ\`a Astronomica Italiana Supplementi 8, 61
%
\bibitem[Sbordone at al.(2010a)]{sbordone10} Sbordone, L., Bonifacio, P., \& Caffau, E.: 2010a, Proceedings of the 11th Symposium on Nuclei in the Cosmos,
Published online at http://pos.sissa.it/cgi-bin/reader/conf.cgi?confid=100, id.294
%
\bibitem[Sbordone et al.(2010b)]{2010A&A...522A..26S} Sbordone, L., Bonifacio, P., Caffau, E., et al.:\ 2010b, A\&A 522, A26 
%
\bibitem[Scannapieco et al.(2009)]{2009MNRAS.396..696S} Scannapieco, C., White, S.~D.~M., Springel, V., \& Tissera, P.~B.:\ 2009, \mnras\ 396, 696 
%
\bibitem[Sch{\"o}nrich \& Binney(2009)]{2009MNRAS.399.1145S} Sch{\"o}nrich, R., \& Binney, J.:\ 2009, \mnras\ 399, 1145 
%
\bibitem[Sch{\"o}rck et al.(2009)]{2009A&A...507..817S} Sch{\"o}rck, T., Christlieb, N., Cohen, J.~G., et al.:\ 2009, A\&A 507, 817 
%
\bibitem[Sharp et al.(2006)]{2006SPIE.6269E..14S} Sharp, R., Saunders, W., Smith, G., et al.:\ 2006, SPIE 6269  
%
\bibitem[Shen et al.(2010)]{2010ApJ...720L..72S} Shen, J., Rich, R.~M., Kormendy, J., et al.:\ 2010, \apjl\ 720, L72 
%
\bibitem[Sneden et al.(2008)]{2008ARA&A..46..241S} Sneden, C., Cowan, J.~J., \& Gallino, R.:\ 2008, \araa\ 46, 241 
%
\bibitem[Spite \& Spite(1982)]{spite82} Spite, M., \& Spite, F.:\ 1982, \nat\ 297, 483
%
\bibitem[Steffen et al.(2009)]{2009MmSAI..80..731S} Steffen, M., Ludwig, H.-G., \& Caffau, E.:\ 2009, Memorie della Societ\`a Astronomica  Italiana\ 80, 73
%
\bibitem[Steinmetz et al.(2006)]{2006AJ....132.1645S} Steinmetz, M., Zwitter, T., Siebert, A., et al.:\ 2006, \aj\ 132, 1645
%
\bibitem[Szentgyorgyi et al.(2011)]{2011PASP..123.1188S} Szentgyorgyi, A., 
Furesz, G., Cheimets, P., et al.\ 2011, \pasp, 123, 1188 
%
\bibitem[Tinsley(1976)]{1976ApJ...208..797T} Tinsley, B.~M.:\ 1976, \apj\ 208, 797 
%
\bibitem[Tolstoy et al.(2009)]{2009ARA&A..47..371T} Tolstoy, E., Hill, V., \& Tosi, M.:\ 2009, \araa\ 47, 371 
%
\bibitem[Tumlinson(2010)]{2010ApJ...708.1398T} Tumlinson, J.:\ 2010, \apj\ 708, 1398 
%
\bibitem[Verschueren et al.(1999)]{1999ASPC..185..108V} Verschueren, W., David, M., \& Vrancken, M.:\ 1999, IAU Colloq.~170: Precise Stellar Radial Velocities 185, 108 
%
\bibitem[Villalobos \& Helmi(2009)]{2009MNRAS.399..166V} Villalobos, {\'A}., \& Helmi, A.:\ 2009, \mnras\ 399, 166 
%
\bibitem[Watson et al.(1998)]{1998SPIE.3355..834W} Watson, F.~G., Parker, 
Q.~A., \& Miziarski, S.\ 1998, \procspie, 3355, 834 
%
\bibitem[Williams et al.(2011)]{2011ApJ...728..102W} Williams, M.~E.~K., Steinmetz, M., Sharma, S., et al.:\ 2011, \apj\ 728, 102 
%
\bibitem[Wilson et al.(2011)]{2011MNRAS.413.2235W} Wilson, M.~L., Helmi, A., Morrison, H.~L., et al.:\ 2011, \mnras\ 413, 2235 
%
\bibitem[Wylie-de Boer \& Cottrell (2009)]{wylie09} Wylie-de Boer, E.~C., \& Cottrell, P.~L.:\ 2009, ApJ 692, 522 
%
\bibitem[Yanny et al.(2009)]{yanny09} Yanny, B., Rockosi, C., Newberg, H.~J., et al.:\ 2009, \aj\ 137, 4377
%
\bibitem[Zoccali et al.(2006)]{2006A&A...457L...1Z} Zoccali, M., Lecureur, A., Barbuy, B., et al.:\ 2006, A\&A 457, L1 
%
\bibitem[Zoccali et al.(2008)]{2008A&A...486..177Z} Zoccali, M., Hill, V., Lecureur, A., et al.:\ 2008, A\&A 486, 177 
%
\end{thebibliography}
\end{document}